\documentclass[a4paper]{jpconf}
\usepackage{graphicx}
\usepackage{amsmath}
\usepackage{amssymb}

\newcommand{\fig}[1]{Figure~\ref{fig:#1}}
\newcommand{\eq}[1]{(\ref{eq:#1})}
\newcommand{\htw}{0.48\textwidth}
\newlength{\pictsize}
\begin{document}
\title{Electrons and Positrons in Cosmic Rays}

\author{A D Panov}

\address{Moscow State University, Skobeltsyn Institute of Nuclear Physics, Moscow, Russia}
\ead{panov@dec1.sinp.msu.ru}


\begin{abstract}
This review concentrates on the results obtained, over the last ten years, on the astrophysics of high-energy cosmic ray electrons and positrons. The anomalies, observed in the data of recent experiments (possible bump in the electron spectrum and the PAMELA anomaly in the positron fraction) are discussed through the systematic use of simple analytical solutions of the transport equations for cosmic ray electrons. Three main ways of explaining the origin of the anomalies are considered: the conservative way supposing the positrons to be pure secondary particles; the nearby sources like pulsars origin; and the dark matter origin. This review discusses, also, the inability to select the pulsars model or the dark matter model to explain the electron anomalies on the basis of the electron spectra with the usual large energy binning ($\gtrsim15\%$). It is argued that the signature of nearby pulsars origin of the anomalies against the dark matter origin could be the fine structure of the cosmic ray electron spectrum predicted in the Malyshev et al. paper (2009) and which was observed in the data from the high-resolution ATIC experiment (2009-2011). 
To date, the high-resolution ATIC data was the only experimental result of this type published in the literature. Therefore, they should be tested by other experiments as soon as possible.
Generally, there is, also, rather controversial situations between the data of the majority of recent experiments and, consequently, there is a noted urgent need for new high-precision and high-statistical experiments.
\end{abstract}


\section{Introduction}

Over the last few years a large amount of new high-precision experimental data related to primary electrons and positrons in cosmic rays was obtained in the experiments of a new generation: PPB-BETS \cite{CRE-EXP-PPB-BETS2008}, ATIC \cite{ATIC-2008-CHANG-NATURE,ATIC-2011-PANOV-ASTRA}, PAMELA \cite{CRE-EXP-POS-PAMELA-NATURE2009,CRE-EXP-PAMELA2010-IzvFIAN,CRE-EXP-PAMELA2011-PhysRevLett}, Fermi/LAT \cite{CRE-EXP-FERMILAT2009A,CRE-EXP-FERMILAT2010B,CRE-EXP-POS-FERMI-2012-PRL}, H.E.S.S. \cite{CRE-EXP-HESS2009}, MAGIC \cite{CRE-EXP-MAGIC2011-arXiv}. 
Some observed features of the data were unexpected and, now compared to the beginning of the 2000's, the situation looks very exciting and intriguing.
The difference can be seen clearly if one compares the content of the present review with D. M{\"u}ller paper \cite{CRE-THEOR-MULLER-2001-ASR} with an almost similar title\footnote{``Cosmic-ray electrons and positrons''.} but published in 2001.
The main subject of present review is the total energy spectrum of cosmic ray electrons and positrons in the energy range higher than approximately 100~GeV, the positron fraction in the total flux, and related astrophysics. 
A number of interesting problems like charge-dependent solar modulation; anisotropy of electron and positron fluxes; and some others are not considered.
Hereinafter, in this paper, the term ``electrons'' is used for the sum of all cosmic ray charged leptons regardless of the charge. Otherwise, ``negative electrons'' or ``positrons'' are used.

\begin{figure}
\begin{minipage}[t]{\htw}
\includegraphics[width=\pictsize]{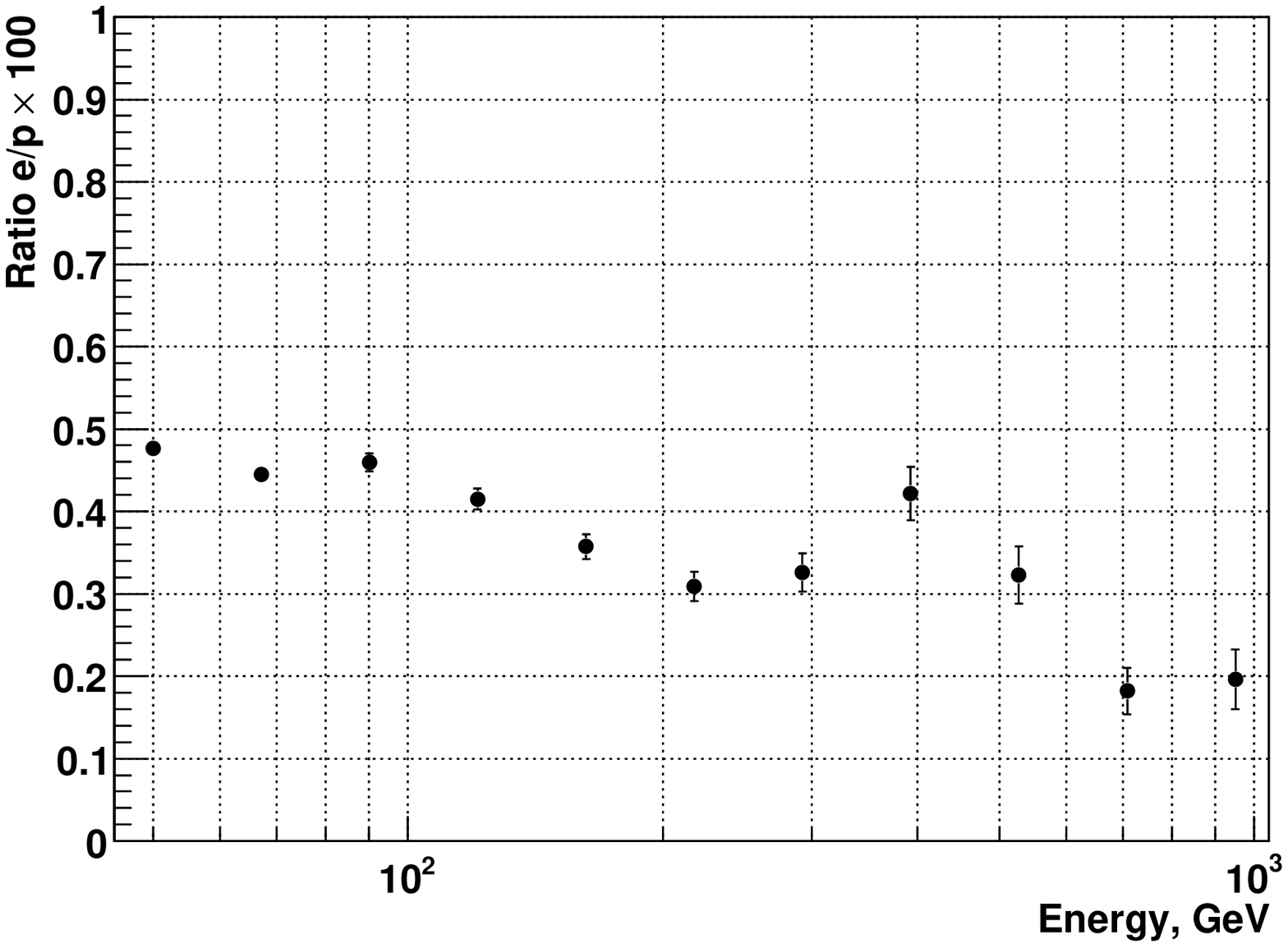}
\caption{\label{fig:RatioElecToProt}
Ratio of the electron flux to the flux of protons given by the ATIC data. The ratio was obtained with the proton spectrum from the paper \cite{ATIC-2009-PANOV-IzvRAN-E} and with the ATIC electron spectrum shown in \fig{AllElectronsCurrentState}, the present paper.
}
\end{minipage}\hfill
\begin{minipage}[t]{\htw}
\includegraphics[width=8cm]{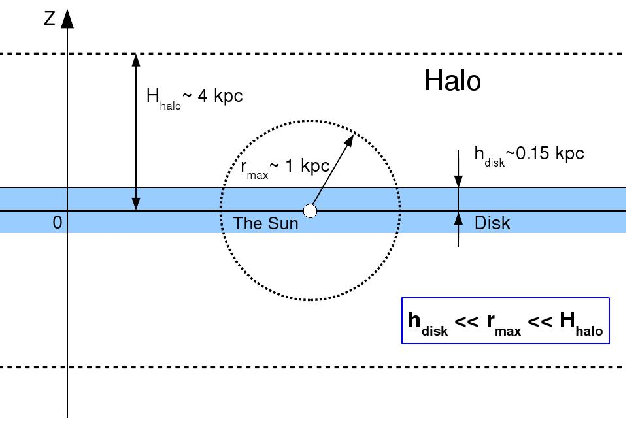}
\caption{\label{fig:Galaxy} 
To the thin disk approximation -- the geometry of the Galactic disk and diffusion halo.
}
\end{minipage}\hspace{2pc}%
\end{figure}

In the total cosmic rays flux, the fraction of electrons is small (see \fig{RatioElecToProt}). However, due to their special properties, they are very important to the astrophysics.
For high-energy electrons ($E \gtrsim 10$~GeV), the dominant channel of energy losses is synchrotron radiation due to interaction of electrons with interstellar magnetic fields and inverse Compton scattering. 
In the Thompson approximation, that holds for electrons very well up to energies around a few TeV, the rate of the energy loss  is given by:
\begin{equation}
   \frac{dE}{dt} = -b_0 E^2 \equiv b(E)
   \label{eq:dEdt}
\end{equation}
where the factor $b_0$ is
\begin{equation}
  b_0 = \frac{4}{3}\frac{\sigma_T c}{(m_e c^2)^2}\left(W_{cmb}+W_{dust}+W_{star}+\frac{B^2}{8\pi}\right).
   \label{eq:bO}
\end{equation}
In \eq{bO} $W_{cmb}$, $W_{dust}$, $W_{star}$ are the energy density of the cosmic microwave background (CMB),
infrared, and starlight photons respectively, $B$ is the mean-square induction of the magnetic field, and $\sigma_T$ is the Thompson cross section. 
Due to the small mass of electrons compared to the proton mass, the radiative energy losses of electrons are much stronger than the radiative losses of protons and nuclei.
The value of $b_0$ for electrons is not known exactly but it is believed widely that $b_0 = (1.2\div1.6)\cdot10^{-16}(\mathrm{GeV}\cdot\mathrm{s})^{-1}$. 


The solution of the equation \eq{dEdt} with the initial condition $E(0) = E_0$ is
\begin{equation}
 E(t) = \frac{E_0}{1+b_0 E_0 t}.
 \label{eq:EtSolution}
\end{equation}
It can be seen easily from \eq{EtSolution} that the electrons with the energy $E_0$ could not be older than 
\begin{equation}
 T_{max}(E_0) = 1/(b_0E_0)
 \label{eq:Tmax}
\end{equation}
since $E(t) \to\infty$ as $t\to -1/(b_0E_0)$. For $E_0=1$~TeV the formula \eq{Tmax} produces the value $T_{max}\sim2\cdot10^5$~years which is much shorter than the lifetime of protons in the Galaxy ($\sim10^7$~years). In the diffusion approximation the mean distance which could be passed by high-energy electrons during their lifetime is determined approximately by the diffusion length
\begin{equation}
 r_{max}\sim\sqrt{2DT_{max}}.
 \label{eq:rmax}
\end{equation}
For the diffusion coefficient $D$, the widely adopted expression is:
\begin{equation}
 D = D_0\cdot(E/\mathrm{GeV})^\delta,
 \label{eq:DE}
\end{equation}
where $\delta = 0.3\div0.6$ and $D_0 = (1\div5)\cdot10^{28}\,\mathrm{sm}^2\mathrm{s}^{-1}$. Then for $b_0 = 1.4\cdot10^{-16}(\mathrm{GeV}\cdot\mathrm{s})^{-1}$ and with formula \eq{DE} , eq.~\eq{rmax} may be written as
\begin{equation}
 r_{max}\sim
  \frac{6.4}{(E/\mathrm{GeV})^{(1-\delta)/2}}\sqrt{\frac{D_0}{3\cdot10^{28}\mathrm{sm}^2\mathrm{s}^{-1}}}\:
  \mathrm{kpc}.
  \label{eq:rmax2}
\end{equation}
For $D_0= 3\cdot10^{28}\,\mathrm{sm}^2\mathrm{s}^{-1}$, one obtains from \eq{rmax2}: $r_{max}(1\:\mathrm{TeV},\delta=0.6)=1.5$~kpc, $r_{max}(1\:\mathrm{TeV},\delta=0.3)=0.6$~kpc. 
It can be seen that the flux of the $\sim$TeV-energy electrons are sensitive to nearby sources in the range of the order of only one kiloparsec and this is very different from the nuclei component of cosmic rays. 
It was recognized clearly more than forty years ago \cite{CRE-THEOR-SHEN1970}. 
The sensitivity of electron flux to a few nearby sources like pulsars or supernova remnants (SNR) makes it very important for astrophysics. 
The other reason of the importance of cosmic ray electrons for astrophysics is a possible connection of them with dark mater (DM) annihilation or decay and with other exotic sources like black hole evaporation etc. 

There are numerous more or less conservative or, on the contrary, very exotic scenarios and models to describe and understand the data related to the cosmic ray electrons. It is convenient to start with the most conservative one -- the so called `conventional model'.


\section{Diffusion approximation, thin disk approximation, and conventional model}

\subsection{The diffusion equation and Green function solution}


Since S.I. Syrovatskii seminal paper \cite{CRE-THEOR-SYROVATSKII1959}, the diffusion approximation is considered to be a reasonable approximation in the transportation problem of cosmic ray electrons.
In the simplest case of homogenious and isotropic diffusion, the diffusion equation reads:
\begin{equation}
 \frac{\partial\rho(\mathbf{r}, E, t)}{\partial t} = \frac{\partial}{\partial E}[b(E)\rho] + D(E)\nabla^2\rho + Q(\mathbf{r}, E, t).
 \label{eq:DiffusionEq}
\end{equation}
Here, $\rho(\mathbf{r}, E, t)$ is the density of electrons of energy $E$; $b(E)$ defined by eq.~\eq{dEdt}; and $Q(\mathbf{r}, E, t)$ is the source function of electrons.
Equation \eq{DiffusionEq} is linear and, therefore, for an arbitrary source function, the solution may be written with Green function $G(\mathbf{r},t,E|\mathbf{r}',t',E')$:
\begin{equation}
 \rho(\mathbf{r},t,E) = 
 \int G(\mathbf{r},t,E|\mathbf{r}',t',E')Q(\mathbf{r}',t',E')d^3\mathbf{r}'dt'dE',
 \label{eq:DiffusionGenSol}
\end{equation}
where the Green function is \cite{CRE-THEOR-SYROVATSKII1959}:
\begin{equation}
 G(\mathbf{r},t,E|\mathbf{r}',t',E')\: =\:
 \dfrac{\exp\left[{\displaystyle-\frac{(\mathbf{r}-\mathbf{r}')^2}{2\lambda^2(E,E')}} \right]\delta(t-t'-\tau(E,E'))}%
 {|b(E)|(2\pi)^{3/2}\lambda^3(E,E')}.
 \label{eq:Green}
\end{equation}
Here
\begin{equation}
 \tau(E,E') = \int_{E}^{E'} \frac{dE''}{|b(E'')|} = \frac{1}{b_0E}-\frac{1}{b_0E'}
\end{equation}
is the time of cooling of an electron from energy $E'$ to $E$ and
\begin{equation}
 \lambda^2(E,E') = 2\int_{E}^{E'} \frac{D(E'')dE''}{|b(E'')|} =
  2D_0(E/\mathrm{GeV})^\delta \left[\frac{1}{b_0E} - \left(\frac{E'}{E}\right)^\delta\frac{1}{b_0E'}\right]
\end{equation}
has the meaning of the mean square of diffusion length during the cooling of an electron from $E'$ to $E$. Hereinafter in this paper we measure $E$ in GeV and do not write explicitly $(E/\mathrm{GeV})$ in formulae.

\subsection{Solution of diffusion equation in thin disk approximation}

In this review we use various simplified analytical solutions of the transport equation \eq{DiffusionEq} systematically to understand and illustrate the physics of cosmic ray electrons.
Whilst such models are not exact, they can be very useful and instructive.
One such model is the model of infinitely thin and homogeneous Galactic disk together with an infinitely thick Galactic halo. 


The half-depth of the diffusion Galatic halo is $\sim4$~kpc which, for 100~GeV--1~TeV electrons, is larger than the expected diffusion electron range $r_{max}\sim1$~kpc. 
At the same time, the half-depth of the Galaxy disk, at the position of the Sun, is only about 150~pc which is much smaller than $r_{max}$ (see \fig{Galaxy}).
Therefore in first approximation, for the sources of electrons located within the Galatic disk, the depth of the halo may be considered as infinitely large.
At the same time, the source of electrons located within the Galatic disk, may be considered to be infinitely thin relative to the value of $r_{max}$. 
Moreover, the Sun is located very close to the Galatic plane, therefore, in this model, the flux of electrons, calculated exactly for $z=0$, is a reasonable approximation in understanding the electron spectrum in the energy range 100~GeV--1~TeV.


Now, we consider a source of electrons with spectrum $Q(E)$, constant in the time, and distributed homogeneously at $z=0$ in the infinitely thin plane. Then, the source function reads:
\begin{equation}
  Q(\mathbf{r},t,E) = Q(E)\delta(z)
  \label{eq:QThin}
\end{equation}
and the solution \eq{DiffusionGenSol} of the transport equation \eq{DiffusionEq} for $z=0$ reduces to:
\begin{equation}
 \rho(E)\rule[-2mm]{0.4pt}{5mm}_{\,z=0} = \frac{1}{b_0\sqrt{2\pi}}\frac{1}{E^2}\int_E^\infty\frac{Q(E')}{\lambda(E,E')}dE'.
  \label{eq:QThinGenSol}
\end{equation}
For power-law spectrum $Q(E)\propto E^{-\gamma}$ and with diffusion coefficient, defined by \eq{DE}, the integral in eq.~\eq{QThinGenSol} is calculated easily and, for the observed spectrum, one obtains:
\begin{equation}
  \rho(E)\rule[-2mm]{0.4pt}{5mm}_{\,z=0} = Q_0 E^{-(\gamma + \Delta)};\quad
  \Delta = \delta + \frac{1}{2},
  \label{eq:QThinPowerLawSol}
\end{equation}
where $Q_0$ is some constant (could be calculated explicitly). Therefore, instead of the source spectrum with the index $\gamma$, an observer measures the electron spectrum steeper on $\Delta = \delta + 1/2$. Since $0.3 < \delta < 0.6$, then $0.8 < \Delta < 1.1$. In other words, $\Delta \approx 1$. It is a rather robust result. It does not depend on the model fine details. 
For example, for a model of infinite homogeneous source, instead of a thin plane source, one checks easily that the solution for power-law spectrum would be $\rho(E) = Q_0 E^{-(\gamma + 1)}$. It is rather close to \eq{QThinPowerLawSol} since $\Delta \approx 1$. Please note that, throughout the paper, we neglect the effects of Solar modulation since these may be important only for the energies below $\sim10$~GeV, however we are interested mainly in energies above 10--100~GeV.

\subsection{``Conventional model'' in the thin disk approximation}
\label{ConvModelThinDisk}


Cosmic ray electron spectrum, which was calculated in I.V. Moskalenko and A.W. Strong paper \cite{CRE-THEOR-SM1998A}, are referred commonly to as ``conventional model''%
\footnote{Also, several variants of similar calculations with various suppositions were published by the same authors later \cite{CRE-THEOR-SM-REIMER2004B,CRE-THEOR-SM-PTUSKIN-2006}.}.
A family of similar models were studied also earlier by R. J. Protheroe in \cite{CRE-THEOR-PROTHEROE1982}. 
The conventional model is the most conservative approach in understanding the cosmic negative electron and positron fluxes. 
In the model, it is supposed  that there exist two main sources of negative electrons and positrons. 
%


Generally, the first kind of sources is the same for primary negative electrons and nuclei (like supernova remnants, SNR). However, sources of this kind do not accelerate positrons.
In the conventional model, this kind of sources is considered in continuous and homogeneous approximation. 
The primary negative electrons source spectral index and the intensity of the source were tuned in \cite{CRE-THEOR-SM1998A} to fit the experimental data\footnote{The most important sources of the data for paper \cite{CRE-THEOR-SM1998A} were papers \cite{CRE-EXP-GOLDEN1984} and \cite{CRE-EXP-NISHIMURA1993}.}.

The second kind of source is secondary electrons produced in inelastic scattering of cosmic ray protons and other nuclei in interstellar medium (ISM) gas. Moreover, it is supposed in the conventional model that this kind of the source is the only source for positrons.


The spectra were calculated in \cite{CRE-THEOR-SM1998A} by numerical solution of diffusion transport equation for a realistic distribution of the substance in the Galaxy. This was done  by using the well-known GALPROP system \cite{STRONG-MOSKALENKO1998-GALPROP} developed by the authors of paper \cite{CRE-THEOR-SM1998A}. 
However it is meaningful to study the conventional model by a more simple analytical solution of the transport equation such as homogeneous thin disk approximation, eqs.~(\ref{eq:QThinGenSol},~\ref{eq:QThinPowerLawSol}) since such models show very clearly the underlying physics and main features of the result.


We suppose power-law source function with spectral index $\gamma_0$ for primary negative electrons and no positrons at all for SNR as a source. The spectral index $\gamma_0$ is not known exactly but it is commonly believed $2.1 < \gamma_0 < 2.5$ and it is expected $\gamma_0$ to be close to the spectral index of proton source spectrum since the sources of the primary electrons and cosmic ray protons are generally the same \cite{CRA-BLANDFORD1980-ApJ,CRE-THEOR-MULLER-2001-ASR}. The index $\gamma_0$ should be considered as a matter of fit of a model to the data. The main source of the secondary electrons is cosmic ray protons that produce them via $\pi^{\pm}$ decay after interaction with ISM nuclei. The spectrum of secondary electrons is a power-law spectrum with the spectral index equal to the spectral index of the observed spectrum of cosmic ray protons, $\gamma_s \approx 2.7$. Then eq.~\eq{QThinPowerLawSol} implies the total spectrum of negative electrons together with positrons to be:

\begin{equation}
 Q_{e^{-}+e^{+}}(E) = Q_0 E^{-(\gamma_0+\Delta)} + Q_{s,e^{-}}E^{-(\gamma_s+\Delta)} + Q_{s,e^{+}}E^{-(\gamma_s+\Delta)},
 \label{eq:ThinGalaxyTotal}
\end{equation}
where $Q_0,Q_{s,e^{-}},Q_{s,e^{+}}$ are amplitudes of the primary negative electron flux, secondary negative electron flux, and secondary positron flux respectively. 
The fit of the conventional model to the data in the paper \cite{CRE-THEOR-SM1998A} shows that the flux of the primary electrons is much higher than the flux of the secondary ones. Therefore the first term in the sum \eq{ThinGalaxyTotal} dominates strongly and, actually, the shape of the spectrum $Q_{e^{-}+e^{+}}(E)$ is close to the power law with the spectral index $(\gamma_0+\Delta)$.

For the fraction of positrons in the total electron flux one obtains easily:
\begin{equation}
 R_{e^{+}/(e^{-}+e^{+})}(E)\equiv
 \dfrac{Q_{e^{+}}(E)}{Q_{e^{-}+e^{+}}(E)} =
 \dfrac{Q_{s,e^{+}}}{Q_{s,e^{-}}+Q_{s,e^{+}}+Q_0 E^{(\gamma_s-\gamma_0)}}.
 \label{eq:PosFracPowModel}
\end{equation}
The value $(\gamma_s-\gamma_0)$ is essentially positive in its nature since it is supposed that primary electrons and protons  have similar source spectrum indexes and $\gamma_s \approx \gamma_0 + \delta$. Therefore, the ratio $R_{e^{+}/(e^{-}+e^{+})}(E)$ is a decreasing function of energy. 


Thus, an approximate power-law spectrum with the spectral index $(\gamma_0+\Delta) \gtrsim 3$ for the total electron flux, and a decreasing fraction of positrons are generic predictions of the conventional model. These features do not depend on details of the conventional model realization.

The conventional model is used commonly as a reference point for discussion of electron data. Also, from the point of view of the conventional model being the first step of the analysis, we present and discuss the data from the experiments.

\section{Electron spectrum and positron fraction: the current state of the experimental studies}

\subsection{The electron spectrum}
\label{ELECRTON-SPECTRUM}

In the total electron spectrum, the most interesting features, measured in recent experiments,  were observed above 100~GeV.
Consequently, we restricted ourselves to reviewing only the data of experiments containing energy points above 100~GeV.
Electron spectra, measured in several well-known but relatively low-energy experiments such as AMS-01 \cite{CRE-EXP-AMS01-2000} and HEAT \cite{CRE-EXP-HEAT2001} were excluded from the review. 


We divided into two classes all the available data on the high-energy cosmic-ray electron spectrum.
We called the first class of data `low-resolution experiments'.
This class related to the spectral data measured with relatively large energy binning for the electron energy: more than approximately 12\%--15\% per bin. 
Even if a spectrometer could provide a higher energy resolution, too large energy binning prevented the observation of details in the spectrum which were finer than the width of the energy bin.
The second class of data addressed measurements with much more narrow energy binning in the electron spectrum together with the high energy resolution of the apparatus.
In such `high-resolution experiments', simultaneously, the energy binning and the resolution were better than 5\%--8\%.
All available data except the only ATIC experiment \cite{ATIC-2011-PANOV-ASTRA} at the time of the preparation of this review, were of the first (low-resolution) class, and the high-resolution class of measurements was represented by the only ATIC measurements \cite{ATIC-2011-PANOV-ASTRA}.
In this section, we review the data from the low-resolution experiments.
The high-resolution experiment \cite{ATIC-2011-PANOV-ASTRA} is discussed separately in Section~\ref{FINESTRUCTURE}.

\begin{figure}
\begin{minipage}[t]{\htw}
\includegraphics[width=\pictsize]{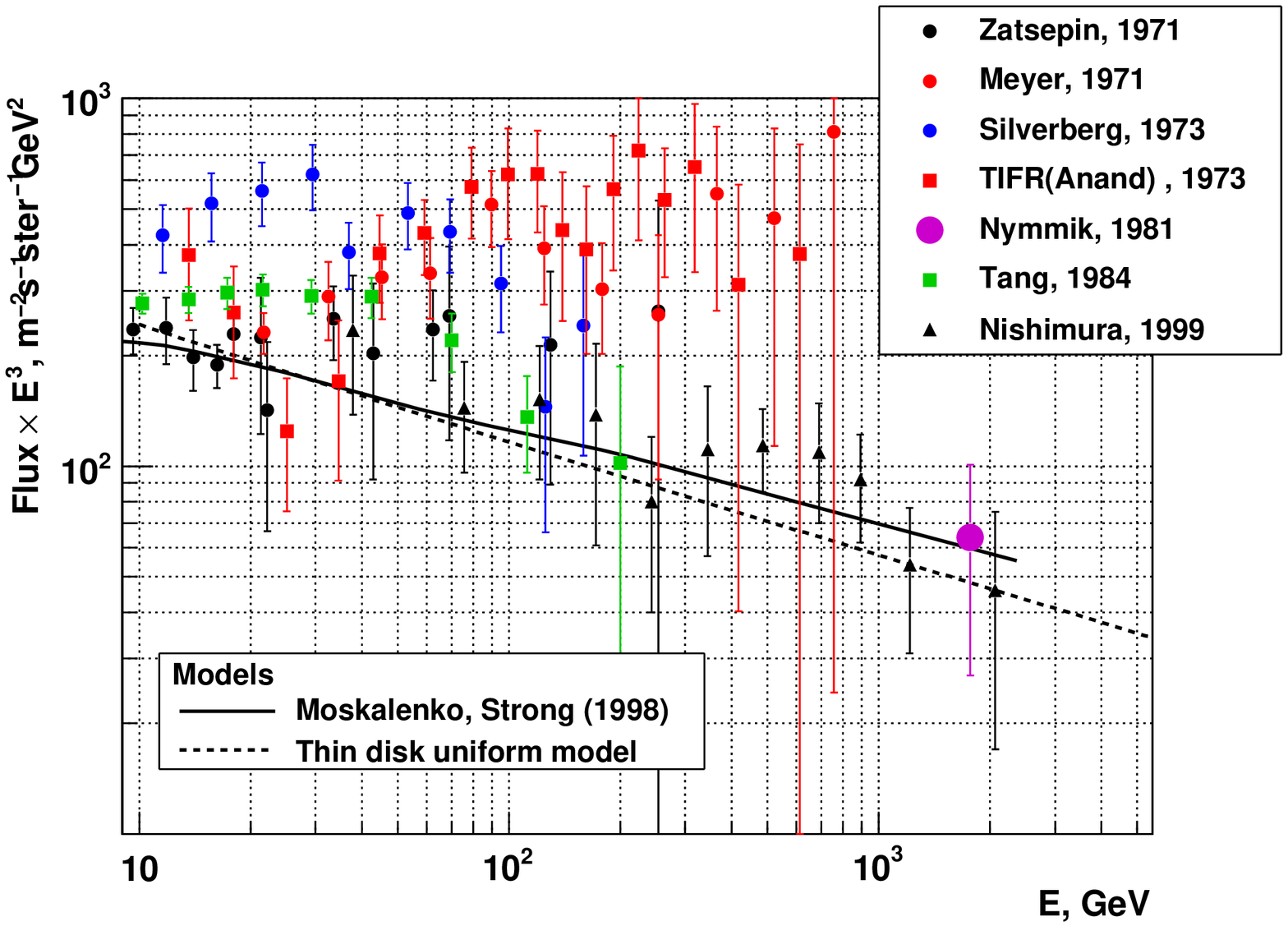}
\caption{\label{fig:AllElectronsBefore2000}
Measuremets of $(e^{+}+e^{-})$-spectrum before 2000 and the conventional model.
The experiments, presented in the figure are: 
Zatsepin, 1971 \cite{CRE-EXP-ZATSEPIN1971};
Meyer, 1971 \cite{CRE-EXP-MEYER1971};
Silberberg, 1973 \cite{CRE-EXP-SILVERBERG1973};
TIFR(Anand), 1973 \cite{CRE-EXP-ANAND1973};
Nymmik, 1981 \cite{CRE-EXP-NYMMIK1981};
Tang, 1984 \cite{CRE-EXP-TANG1984};
Nishimura, 1999 \cite{CRE-EXP-NISHIMURA1999} (see also earlier papers of the same experiment
\cite{CRE-EXP-NISHIMURA1979,CRE-EXP-NISHIMURA1980,CRE-EXP-NISHIMURA1993}).
}
\end{minipage}\hfill
\begin{minipage}[t]{\htw}
\includegraphics[width=\pictsize]{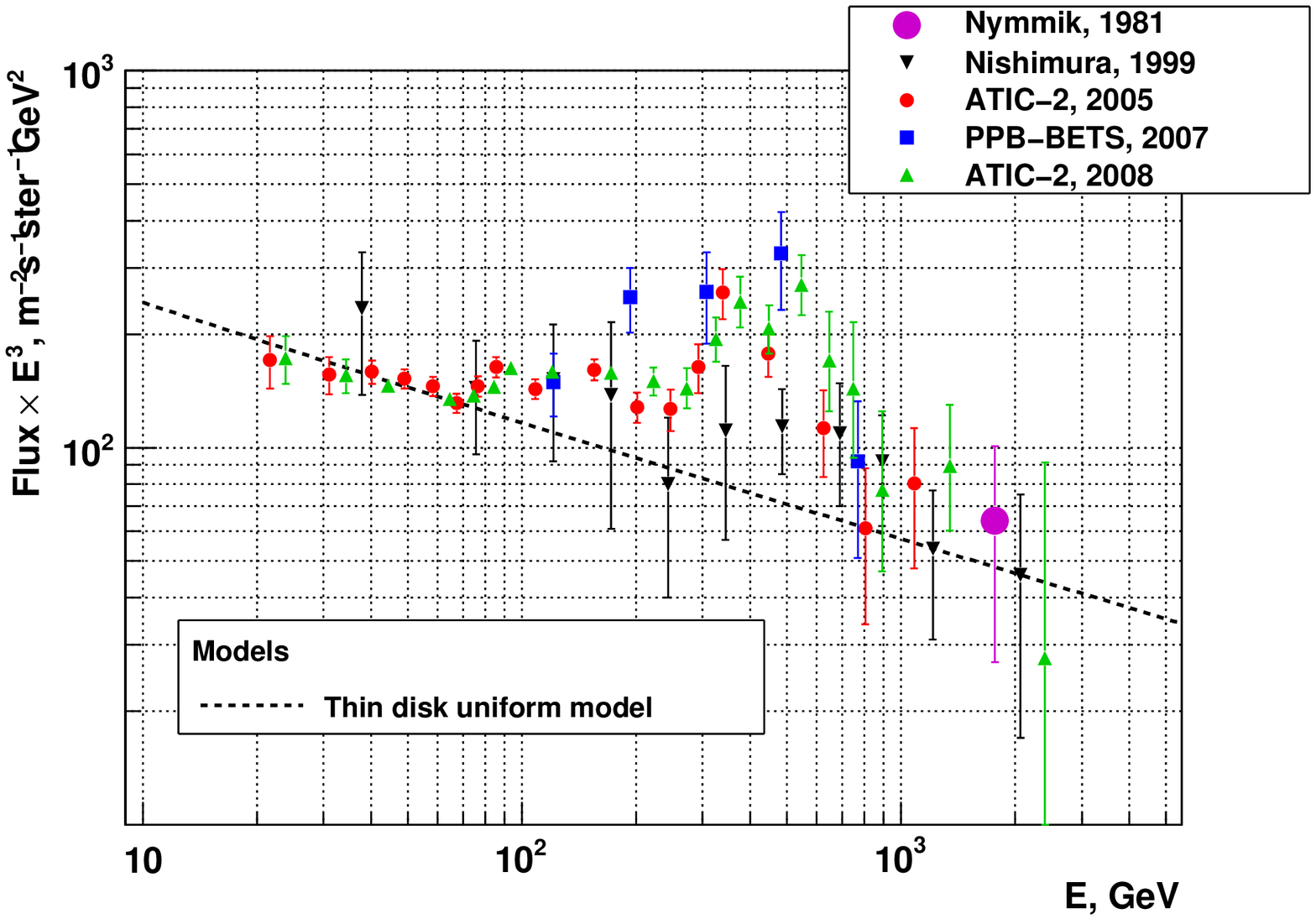}
\caption{\label{fig:AllElectronsATIC2005}
Measuremets of $(e^{+}+e^{-})$-spectrum, 2005--2008 and some older data: Nymmik, 1981 \cite{CRE-EXP-NYMMIK1981}, Nishimura, 1999 \cite{CRE-EXP-NISHIMURA1999}, ATIC-2, 2005 \cite{ATIC-2005-CHANG-ICRC}, PPB-BETS, 2007 \cite{CRE-EXP-PPB-BETS2008-ICRC,CRE-EXP-PPB-BETS2008}, ATIC-2, 2008 \cite{ATIC-2008-CHANG-NATURE}.
}
\end{minipage}\hspace{2pc}\\%
\end{figure}
%


\fig{AllElectronsBefore2000} shows a compilation of the electron spectrum data obtained before 2000.
The experiments were of a different nature: these were both balloon-borne \cite{CRE-EXP-ZATSEPIN1971,CRE-EXP-MEYER1971,CRE-EXP-SILVERBERG1973,CRE-EXP-ANAND1973,CRE-EXP-TANG1984,CRE-EXP-NISHIMURA1999} and satellite \cite{CRE-EXP-NYMMIK1981}. 
They used different techniques including emulsion chambers \cite{CRE-EXP-ZATSEPIN1971,CRE-EXP-ANAND1973,CRE-EXP-NYMMIK1981}, \v{C}erenkov counters/calorimetric \cite{CRE-EXP-MEYER1971,CRE-EXP-SILVERBERG1973}, transition radiation/calorimetric \cite{CRE-EXP-TANG1984}. 
A review of experiments before 2000 may be found also in \cite{CRE-THEOR-MULLER-2001-ASR}. 
The Moskalenko and Strong conventional model \cite{CRE-THEOR-SM1998A} and the uniform thin disk model, described in the Section \ref{ConvModelThinDisk}, eq.~\eq{ThinGalaxyTotal}, are shown, also, in \fig{AllElectronsBefore2000} together with the data.
The parameters of the thin disk model were $\gamma_0=2.3$, $\gamma_s=2.7$, $\Delta=1.0$, $Q_0=450\,\mathrm{m}^{-2}\mathrm{sr}^{-1}\mathrm{sec}^{-1}$, $Q_{s,e^{-}}=24\,\mathrm{m}^{-2}\mathrm{sr}^{-1}\mathrm{sec}^{-1}$, $Q_{s,e^{+}}=62\,\mathrm{m}^{-2}\mathrm{sr}^{-1}\mathrm{sec}^{-1}$. Here $Q_{s,e^{+}}/Q_{s,e^{-}} \approx 2.6$ was the ratio of positrons to negative electrons production rates in p-p collisions \cite{CRE-THEOR-SM1998A}.


The situation at the beginning of the 2000's looked very controversial.
There were two different groups of data. The first group suggested a soft electron spectrum ($\gamma > 3$) at energies above $\sim50$~GeV: Silverberg, 1973 \cite{CRE-EXP-SILVERBERG1973}; Nymmik, 1981 \cite{CRE-EXP-NYMMIK1981}; Tang, 1984 \cite{CRE-EXP-TANG1984}; and Nishimura, 1999 \cite{CRE-EXP-NISHIMURA1999} (see \fig{AllElectronsBefore2000}). 
The second group suggested a hard electron spectrum ($\gamma < 3$) and a much higher intensity of the spectrum above 100~GeV: Meyer 1971, \cite{CRE-EXP-MEYER1971} and TIFR(Anand), 1973 \cite{CRE-EXP-ANAND1973}. 
The behavior of the data in the experiment Zatsepin, 1971 \cite{CRE-EXP-ZATSEPIN1971} was somewhat median between these two groups. 
The statistics of the experiments were low and it was impossible to discuss seriously the existence of structures like bumps in the spectra.
It was difficult to obtain some definite conclusions on the basis of the collection of the data completed before 2000.


%
\begin{figure}
\begin{minipage}[t]{\htw}
\includegraphics[width=\pictsize]{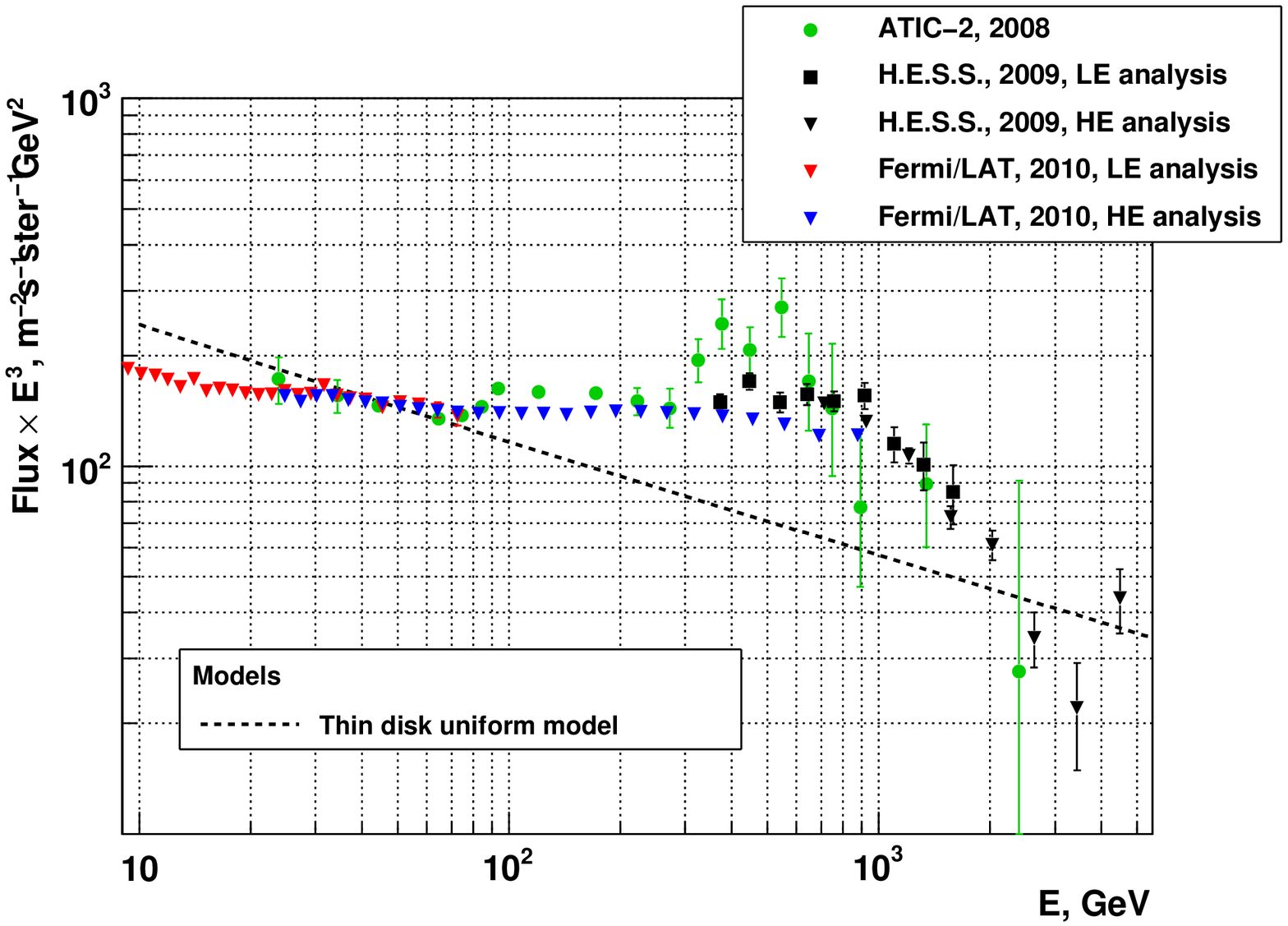}
\caption{\label{fig:AllElectronsFermiHESS}
Measuremets of $(e^{+}+e^{-})$-spectrum: ATIC-2, 2008 \cite{ATIC-2008-CHANG-NATURE}, H.E.S.S., 2009 \cite{CRE-EXP-HESS2009}, Fermi/LAT, 2010 \cite{CRE-EXP-FERMILAT2010B}.
}
\end{minipage}\hfill
\begin{minipage}[t]{\htw}
\includegraphics[width=\pictsize]{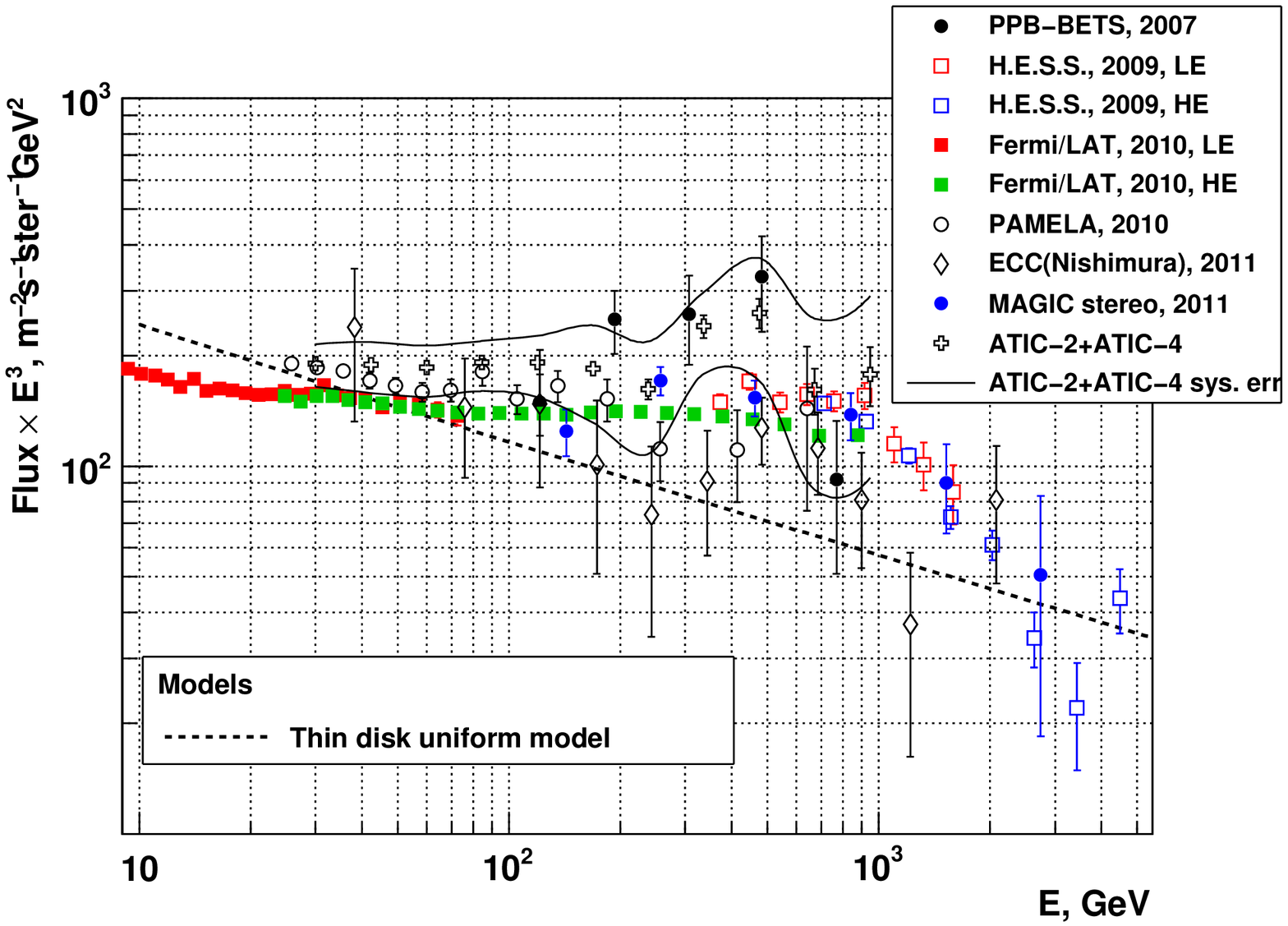}
\caption{\label{fig:AllElectronsCurrentState}
Measuremets of $(e^{+}+e^{-})$-spectrum, the current state of the art: PPB-BETS, 2007 \cite{CRE-EXP-PPB-BETS2008-ICRC,CRE-EXP-PPB-BETS2008}; H.E.S.S., 2009 \cite{CRE-EXP-HESS2009}; Fermi/LAT, 2010 \cite{CRE-EXP-FERMILAT2010B}; PAMELA, 2010 \cite{CRE-EXP-PAMELA2010-IzvFIAN}; ECC(Nishimura), 2011 \cite{CRE-EXP-NISHIMURA2011-ICRC}; MAGIC stereo, 2011 \cite{CRE-EXP-MAGIC2011-arXiv}; and ATIC2$+$ATIC-4 -- the spectrum of the paper \cite{ATIC-2011-PANOV-ASTRA} shown with enlarged energy bins.
}
\end{minipage}%
\end{figure}


There were no new data on high-energy cosmic ray electron spectrum from 2000 to 2004, but in 2005 the preliminary electron spectrum measured by the Antarctic long-duration balloon-born experiment ATIC-2 (the second flight of the ATIC spectrometer) were published \cite{ATIC-2005-CHANG-ICRC} (see \fig{AllElectronsATIC2005}). 
ATIC covered the energy range from 30~GeV to more than 2~TeV and provided the data with much better statistics than from all previous experiments. There were a number of new features in the data.
The electron spectrum showed that the spectral index was close to 3.0 in the energy range 30--300~GeV; and there was a hint on energy cut-off near 1 TeV (however, with low statistical confidence) and, more importantly, a prominent bump between 300 and 600~GeV.
Thise behaviour disagreed with the conventional model; however, between 2005 to 2007, the phenomenon was not discussed seriously.


In 2007, the data  from the Antarctic PPB-BETS experiment  was presented  at the 30th International Cosmic Ray Conference and published in 2008 \cite{CRE-EXP-PPB-BETS2008,CRE-EXP-PPB-BETS2008-ICRC} (see \fig{AllElectronsATIC2005}).
The PPB-BETS spectrum demonstrated, also, a bump-like structure between 100 and 700~GeV which was contradictory to the expectation of the conventional model.
The statistics were lower than in previous ATIC experiment and the shape of the bump was somewhat different than in ATIC data. 
Near the end of 2008, the ATIC paper on the electron spectrum was published in Nature \cite{ATIC-2008-CHANG-NATURE}.
The bump, in the preliminary spectrum \cite{ATIC-2005-CHANG-ICRC}, was confirmed and all this provoked a wide discussion of the nature of the phenomenon in the literature (see below).
The bumps, observed by ATIC and PPB-BETS, looked similar to the bump observed, also, in 1999 in the spectrum measured in the Japanese emulsion experiment \cite{CRE-EXP-NISHIMURA1999} (however, with much lower statistics, see \fig{AllElectronsATIC2005}).  Consequently, the existence of this bump-like feature in the electron spectrum proved rather more conclusive this time (2008).

The situation became less clear after the publication, in 2009,  of the electron spectrum measured by the space spectrometer Fermi/LAT \cite{CRE-EXP-FERMILAT2009A} and then in 2010 \cite{CRE-EXP-FERMILAT2010B} (see \fig{AllElectronsFermiHESS}).
Fermi/LAT provided very high statistics and confirmed ATIC data in the energy range 30--200~GeV by showing the spectral index to be close to 3.0.
However, at higher energies, the Fermi/LAT spectrum did not exhibit any prominent spectral features similar to the ATIC bump.
The Fermi/LAT instrument had a lower energy resolution than the ATIC spectrometer due to the thin calorimeter of Fermi/LAT against the ATIC thick for electrons calorimeter. However, this reason could not explain the lack of the structures in the Fermi/LAT data.
Also, in 2009, a  complete analysis (low-energy analysis together with high-energy analysis) of the electron spectrum, measured by a ground-based \v{C}erenkov telescope H.E.S.S. in the energy range 350~GeV--5~TeV, was published in the paper \cite{CRE-EXP-HESS2009} (the high-energy part of the spectrum was published previously in 2008 \cite{CRE-EXP-HESS2008}).
The H.E.S.S. telescope data confirmed, with better statistics, the high-energy cut-off in the electron spectrum measured by ATIC. However, due to the high energy threshold of the telescope, it  could not confirm or disprove clearly the bump in the energy range 250-700 GeV (\fig{AllElectronsFermiHESS}).


New measurements of cosmic ray electron spectrum were carried out in 2010--2012 after Fermi/LAT and H.E.S.S. measurements (\fig{AllElectronsCurrentState}). 
The PAMELA space spectrometer electron spectrum, measured in the energy range 25--640~GeV, was published in 2010 \cite{CRE-EXP-PAMELA2010-IzvFIAN}.
In the energy range 25--200~GeV, the spectrum was similar to the spectra measured by ATIC and Fermi/LAT and showed the spectral index to be close to 3.0. However, at energy near 200~GeV, the PAMELA spectrum showed some down step-like features.
For energies above 200~GeV, it provided an absolute flux lower than the ATIC and Fermi/LAT spectra. However, the statistics of the data were low and this conclusion was not firm.
The PAMELA collaboration published, also, (in two different versions), the spectrum of negative electrons \cite{CRE-EXP-PAMELA2011-PhysRevLett}. However, it would be incorrect to compare it with the total $e^{-} + e^{+}$ spectra measured by other experiments discussed here.
The collaboration of Japanese emulsion experiments \cite{CRE-EXP-NISHIMURA1979,CRE-EXP-NISHIMURA1980,CRE-EXP-NISHIMURA1993,CRE-EXP-NISHIMURA1999} published, in 2011 \cite{CRE-EXP-NISHIMURA2011-ICRC}, the electron spectrum with updated and improved statistics.
ECC was the new name of the experiment. This ECC spectrum confirmed well the previous version of the data \cite{CRE-EXP-NISHIMURA1999}.
The electron spectrum was measured, also, by the new MAGIC ground-based \v{C}erenkov stereo telescope.
The first preliminary version of the spectrum was published in the dissertation \cite{CRE-EXP-MAGIC2012-BorlaThes} and, then, improved data was published in the paper \cite{CRE-EXP-MAGIC2011-arXiv} (shown in \fig{AllElectronsCurrentState}).
The spectrum showed a weak bump-like feature with the maximum near 250~GeV; however, the statistical significance of the structure  was low.
Finally, the electron spectrum, measured by the the ATIC spectrometer $4^\mathrm{th}$ flight was obtained and published together with updated data of the ATIC-2 \cite{ATIC-2011-PANOV-ASTRA} spectrum.
Generally, the ATIC-2$+$ATIC-4 spectrum (see \fig{AllElectronsCurrentState}) confirmed previous ATIC-2 \cite{ATIC-2008-CHANG-NATURE} data with minor corrections.
The paper \cite{ATIC-2008-CHANG-NATURE} main purpose was a measurement of high-resolution spectrum of electrons (see Section~\ref{FINESTRUCTURE}). However,  for convenience and in order to compare it with other data, the spectrum, measured in \cite{ATIC-2008-CHANG-NATURE}, is shown in \fig{AllElectronsCurrentState} with enlarged energy bins.


The current situation with experimental electron spectrum looks very controversial. 
There are a number of modern experiments  which provide a bump-like structure in the energy range 250--700~GeV -- ECC; PPB-BETS; ATIC; and, might be, MAGIC. However, in this region, the Fermi/LAT spectrometer showed no bump-like feature.
It is important that the Fermi/LAT experiment is a very high-statistical one; however, the statistical significance of the data from PPB-BETS and ATIC experiments also was rather good. 
Obviously, in some experiments, there were unaccounted systematic errors and, even, the expected reported systematics were generally large (for example, see the systematic error corridor for the ATIC-2$+$ATIC-4 spectrum in \fig{AllElectronsCurrentState}).
We can say, with confidence, that now, the main problem is not the lack of statistics but unaccounted systematic errors in the experiments.


Whilst the situation in relation to the existence of some bump-like structure in the electron spectrum is completely unclear, the data from all high-statistical measurements are in clear contradiction to the conventional model prediction: the spectrum is harder than the expected one and has a cutoff near 1~TeV.


\begin{figure}
\begin{minipage}[t]{\htw}
\includegraphics[width=\pictsize]{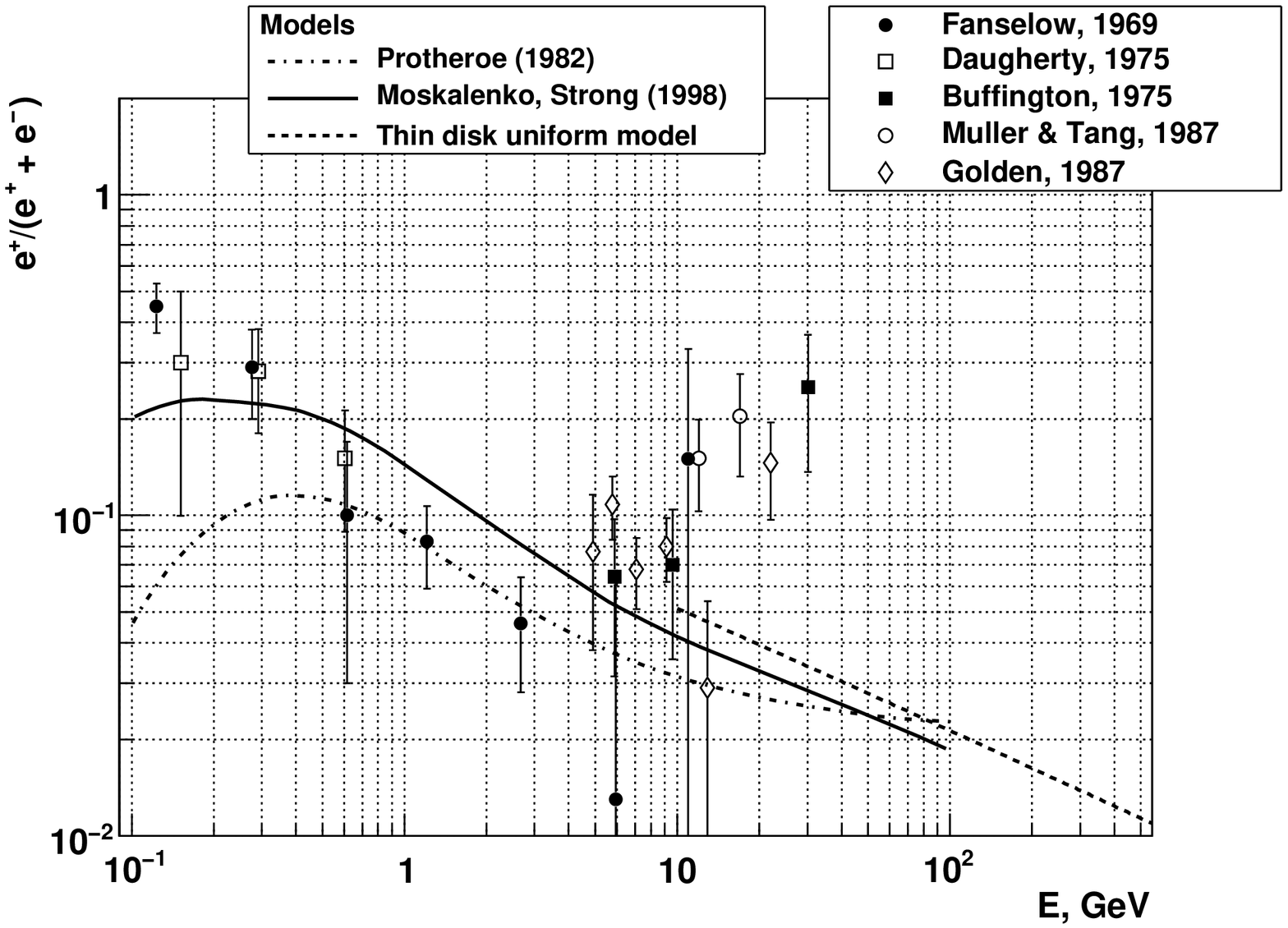}
\caption{\label{fig:PositronsFractionBefore1990}
Measuremets of positron fraction $e^{+}/(e^{+}+e^{-})$ before 1990: Fanselow, 1969 \cite{CRE-EXP-POS-FANSELOW1969}; Daugherty, 1975 \cite{CRE-EXP-POS-DAUGHERTY1975}; Buffington, 1975 \cite{CRE-EXP-POS-BUFFINGTON1975}; Muller \& Tang, 1987 \cite{CRE-EXP-POS-MULLER1987}; Golden, 1987 \cite{CRE-EXP-POS-GOLDEN1987}. Models: Protheroe (1982) \cite{CRE-THEOR-PROTHEROE1982}; Moskalenko, Strong (1998) \cite{CRE-THEOR-SM1998A}; Thin disk uniform model -- this work.
}
\end{minipage}\hfill
\begin{minipage}[t]{\htw}
\includegraphics[width=\pictsize]{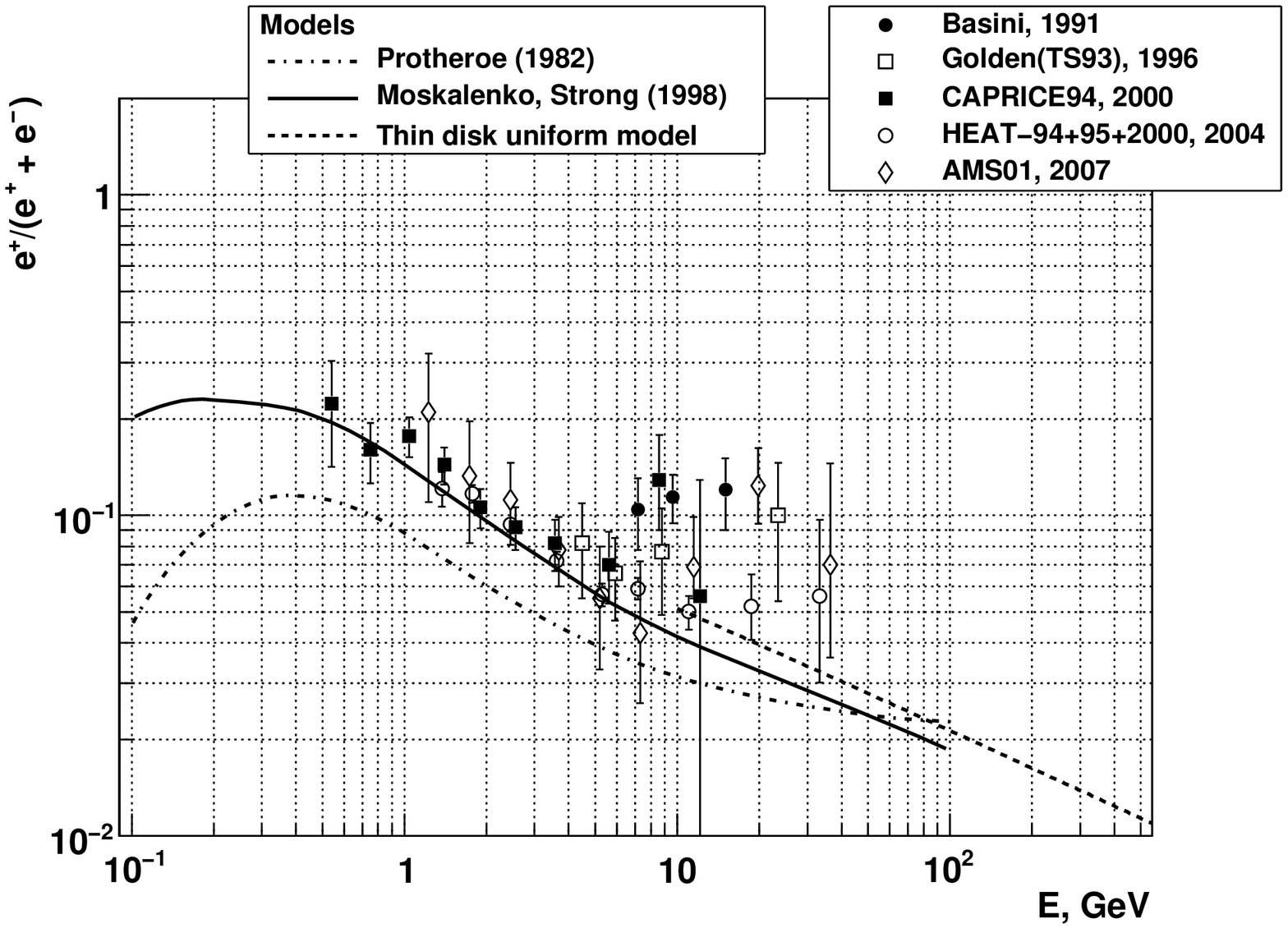}
\caption{\label{fig:PositronsFractionBefore2008}
Measuremets of positron fraction $e^{+}/(e^{+}+e^{-})$ before PAMELA (2008): Basini, 1991 \cite{CRE-EXP-POS-BASINI1991}; Golden(TS93), 1996 \cite{CRE-EXP-POS-GOLDEN1996}; CAPRICE94, 2000 \cite{CRE-EXP-POS-CAPRICE2000}; HEAT-94$+$05$+$2000, \cite{CRE-EXP-POS-HEAT2004}; AMS01, 2007 \cite{CRE-EXP-POS-AMS01-2007}.
}
\end{minipage}\\
\begin{minipage}[t]{\htw}
\includegraphics[width=\pictsize]{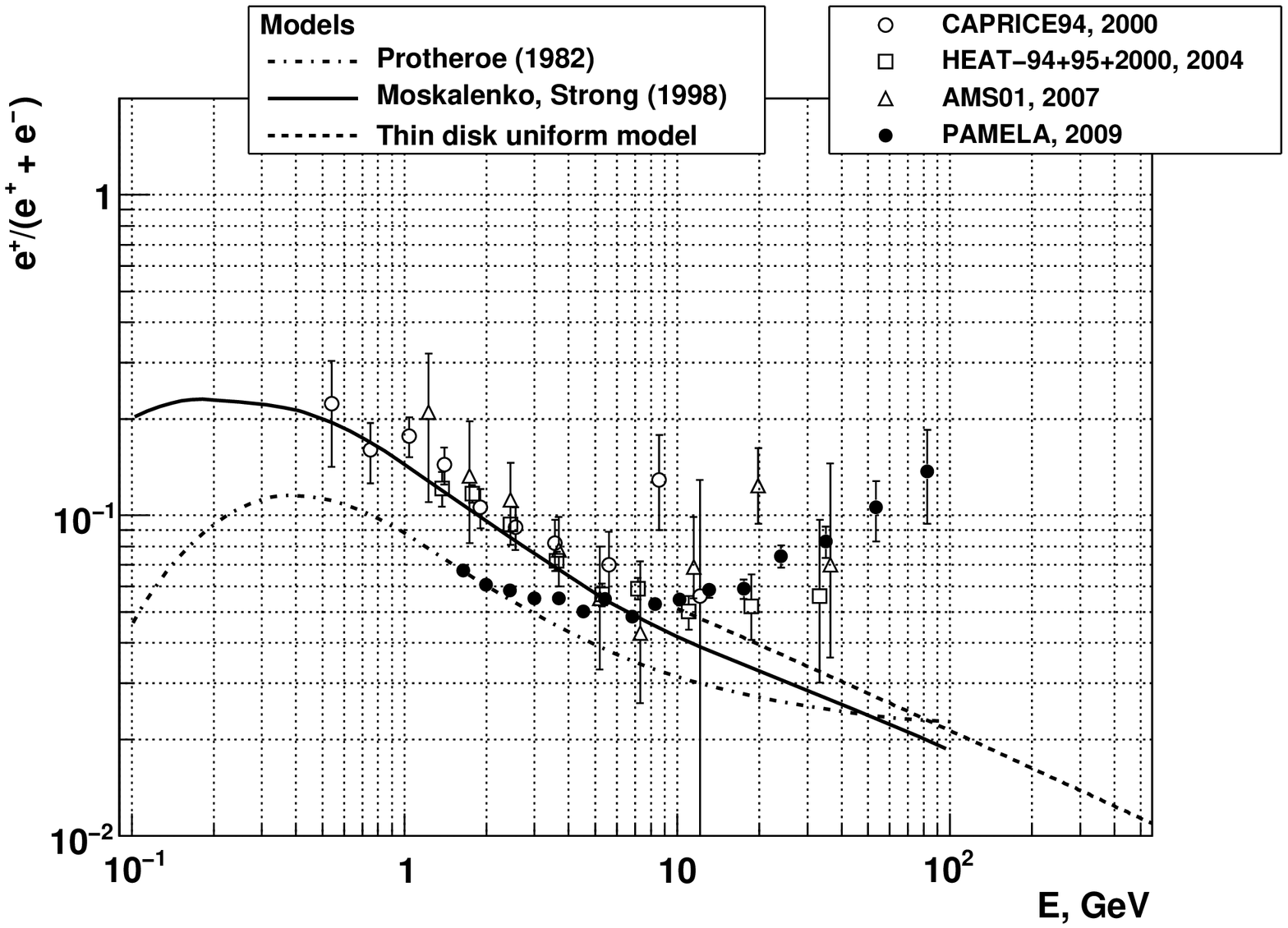}
\caption{\label{fig:PositronsFractionPAMELA}
Measuremets of positron fraction $e^{+}/(e^{+}+e^{-})$: the first PAMELA data and some previous measurements: CAPRICE, 2000 \cite{CRE-EXP-POS-CAPRICE2000}; HEAT-94$+$05$+$2000, 2004 \cite{CRE-EXP-POS-HEAT2004}; AMS01, 2007 \cite{CRE-EXP-POS-AMS01-2007}; PAMELA, 2009 \cite{CRE-EXP-POS-PAMELA-NATURE2009}.
}
\end{minipage}\hfill
\begin{minipage}[t]{\htw}
\includegraphics[width=\pictsize]{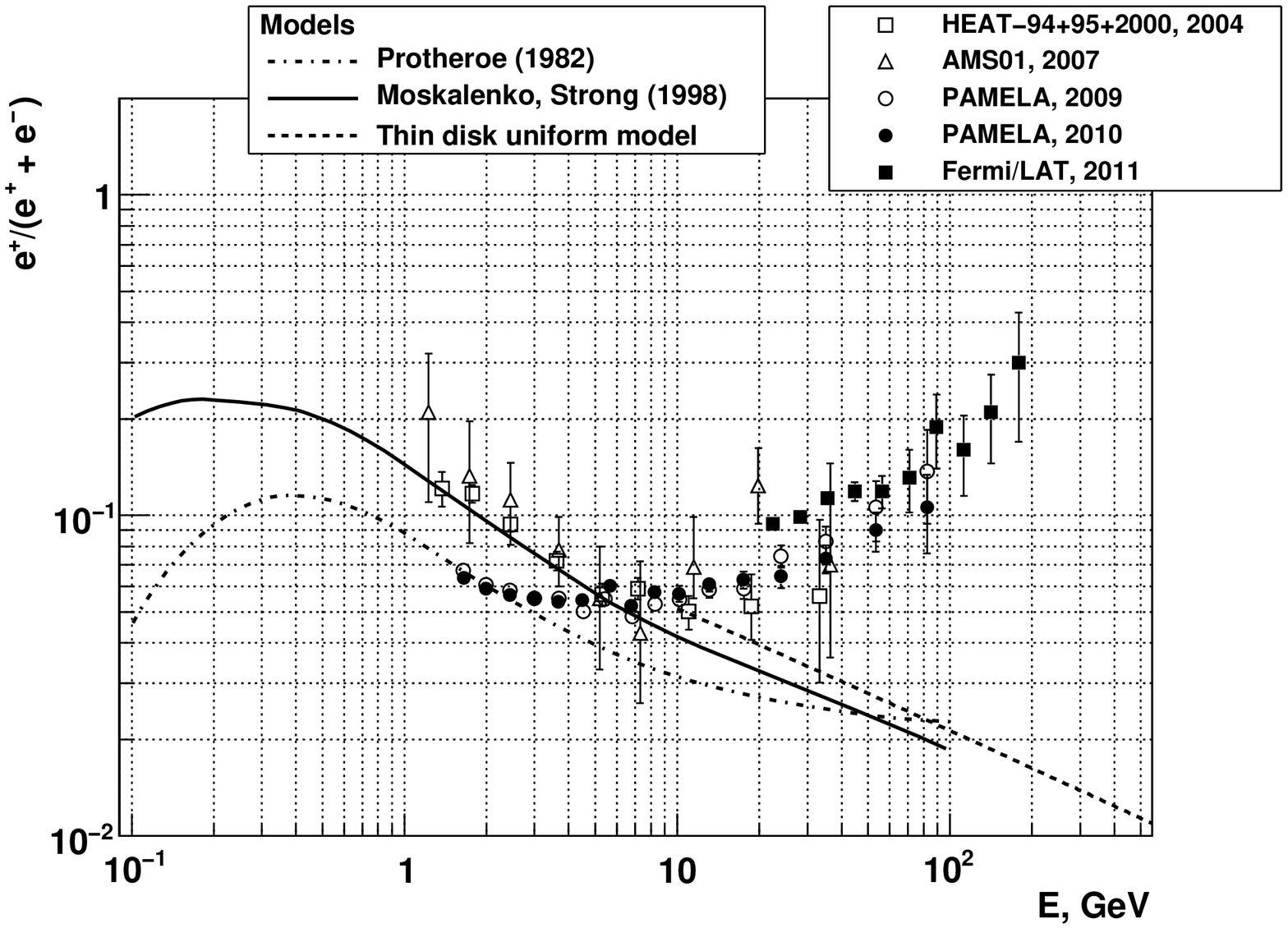}
\caption{\label{fig:PositronsFractionContemp}
Measuremets of positron fraction $e^{+}/(e^{+}+e^{-})$,  all data after 2000:  HEAT-94$+$05$+$2000, 2004 \cite{CRE-EXP-POS-HEAT2004}; AMS01, 2007 \cite{CRE-EXP-POS-AMS01-2007}; PAMELA, 2009 \cite{CRE-EXP-POS-PAMELA-NATURE2009}; PAMELA, 2010 \cite{CRE-EXP-POS-PAMELA-2011-APh}; Fermi/LAT, 2011 \cite{CRE-EXP-POS-FERMI-2012-PRL}.
}
\end{minipage}\hfill
\end{figure}


\subsection{The positron fraction}


\fig{PositronsFractionBefore1990} shows the data on positron fraction in the total $e^{+}+e^{-}$ flux, obtained before 1990, together with  the three conventional-type models.
It can be seen in \fig{PositronsFractionBefore1990}, that already the data of experiments in the 1970's and 1980's, showed  a deviation above 10~GeV from the simple conventional models.
There was an increase of the positron fraction instead of a reduction predicted by the conventional models (see eq.~\eq{PosFracPowModel}).
At high energies, the deviation of the data from the expectations was recognized clearly and, in the 1980's,  noted in a number of papers \cite{CRE-THEOR-PROTHEROE1982,CRE-EXP-POS-MULLER1987}. However, the statistics were too low to reach firm conclusions about it. 


\fig{PositronsFractionBefore2008} shows the data, obtained between 1990 and 2008 (before the PAMELA experiment).
Generally, these confirmed a deviation in the positron fraction from the conventional model predictions. However, the statistics were still too low and the deviation was not discussed widely in the literature (however, it was noted again, see  \cite{CRE-THEOR-AHARONIAN1995-A+A,CRE-THEOR-ATOYAN1995-PhysRevD,CRE-THEOR-ZHANG2001,CRE-THEOR-KANE2002}).


The situation changed dramatically after the publication, in 2009 \cite{CRE-EXP-POS-PAMELA-NATURE2009} (\fig{PositronsFractionPAMELA}), of the data from the space spectrometer PAMELA.
These high-statistical and high-precision data from the space magnetic spectrometer PAMELA showed definitely that the positron fraction increased at energies above 10~GeV.
The positron fraction, measured by PAMELA as below 5~GeV, was lower than the data from previous experiments \cite{CRE-EXP-POS-CAPRICE2000,CRE-EXP-POS-HEAT2004,CRE-EXP-POS-AMS01-2007} which provided, also, a sufficiently high precision in this region.
This difference was not a contradiction and it could be explained by different charge-sign dependent solar modulations at different epochs of solar magnetic polarity.
Such modulation variations were confirmed directly in a series of AESOP measurements \cite{CRE-EXP-POS-CLEM2008}.


All the modern measurements of the positron fraction after 2000 are shown in \fig{PositronsFractionContemp}.
Recently, the PAMELA \cite{CRE-EXP-POS-PAMELA-NATURE2009} 2009 data was confirmed again by the PAMELA spectrometer through new experimental data and with the application of three novel proton background models \cite{CRE-EXP-POS-PAMELA-2011-APh}.
\fig{PositronsFractionContemp} shows the data related to beta-fit (fit of background with using of beta-function, see details in \cite{CRE-EXP-POS-PAMELA-2011-APh}).
Finally, the space spectrometer Fermi/LAT measured indirectly the positron fraction by using the Earth magnetic field to separate negative electrons from positrons \cite{CRE-EXP-POS-FERMI-2012-PRL}.
The increasing positron fraction was confirmed up to the energies about 200~GeV, and the positron fraction achieved the value $\sim0.3$ (see \fig{PositronsFractionContemp}).


Thus, there are two puzzles related to electrons in the cosmic rays: no power-law spectrum of electrons (possibly with a bump) (\fig{AllElectronsCurrentState}) and the increase in the fraction of positrons above 10~GeV (\fig{PositronsFractionContemp}).
Both phenomena contradict the conventional model. Therefore, probably, either the conventional model is oversimplified or incorrect in some respects. The cause of these deviations of the data from the model should be understood.


\section{Three main ways to explain the anomalies in the electron spectrum and positron fraction}
\label{THREE-WAYS}


The publication of the electron spectrum, measured by ATIC \cite{ATIC-2008-CHANG-NATURE} and the positron fraction, measured by PAMELA \cite{CRE-EXP-POS-PAMELA-NATURE2009}, resulted in a burst of theoretical thoughts which started from 2008--2009.
Hundreds of papers with discussions and explanations of the data were published.
It is impossible to review all the related papers here.
Instead, we consider and classify here, with a number of examples, only three main directions of theoretical investigations. 


\subsection{Conservative models}


Under the conservative way, we mean those models which try to save the conventional model main supposition which is to consider cosmic ray positrons as pure secondary particles.
Generally, in such models, it is possible to explain bumps and energy cutoff in the total electron spectrum with sources like local SNRs or an inhomogeneity of the sources distribution \cite{CRE-EXP-NISHIMURA1997,CRE-THEOR-EW2002,CRE-THEOR-EW2009}. However, the main problem is the PAMELA anomaly in the positron fraction.
There are only a few papers, following the ATIC \cite{ATIC-2008-CHANG-NATURE} and PAMELA \cite{CRE-EXP-POS-PAMELA-NATURE2009} results, which try to understand them within the conservative model.

\begin{figure}
\begin{minipage}[t]{\htw}
\includegraphics[width=\pictsize]{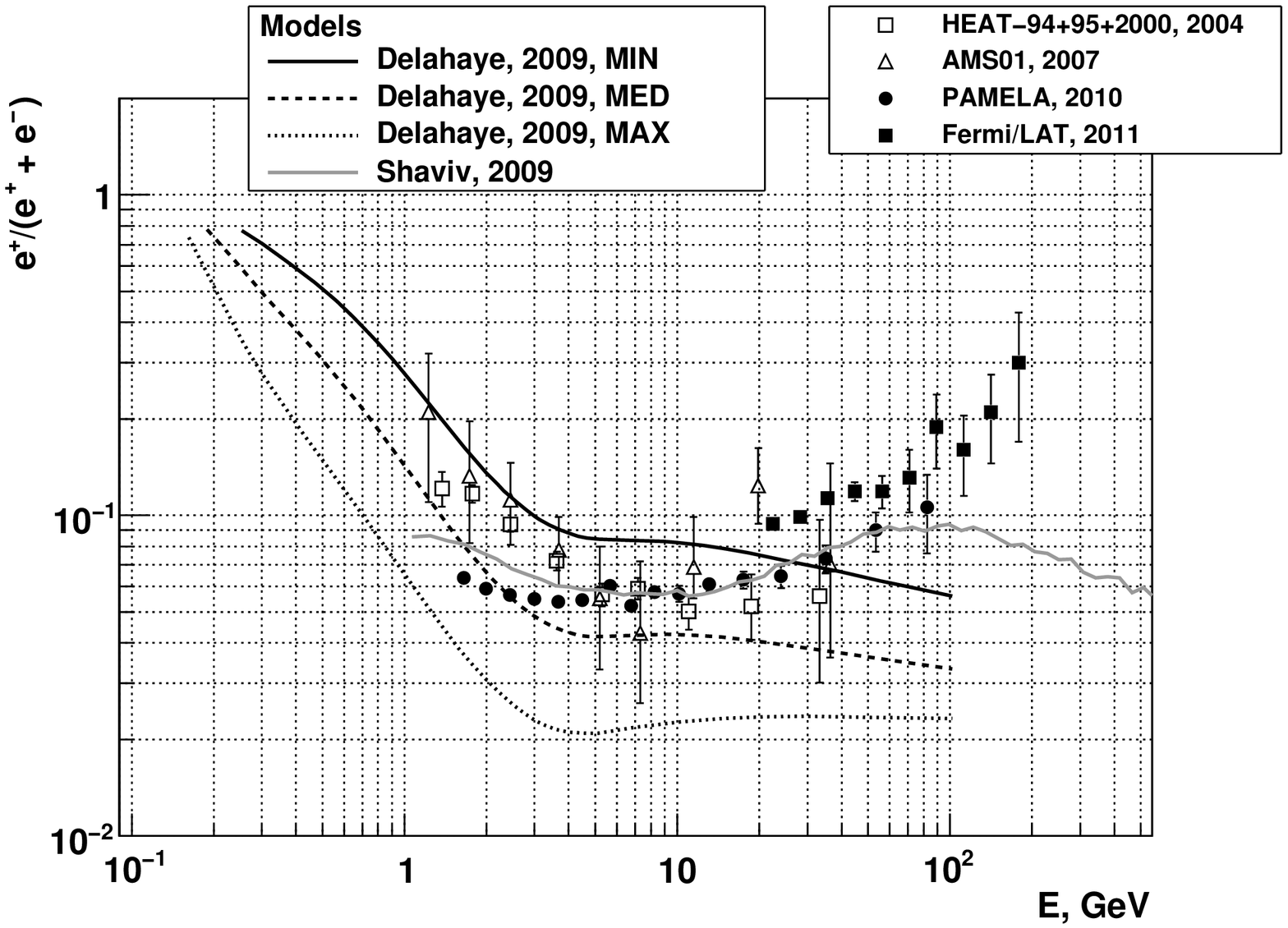}
\caption{\label{fig:PositronsFraction-DelahayShaviv}
Positron fraction $e^{+}/(e^{+}+e^{-})$: conservative models of T.~Delahaye et al and N.J.~Shaviv et al , and  all data after 2000:  HEAT-94$+$05$+$2000, 2004 \cite{CRE-EXP-POS-HEAT2004}; AMS01, 2007 \cite{CRE-EXP-POS-AMS01-2007}; PAMELA, 2009 \cite{CRE-EXP-POS-PAMELA-NATURE2009}; PAMELA, 2010 \cite{CRE-EXP-POS-PAMELA-2011-APh}; Fermi/LAT, 2011 \cite{CRE-EXP-POS-FERMI-2012-PRL}.
}
\end{minipage}\hfill
\begin{minipage}[t]{\htw}
\includegraphics[width=\pictsize]{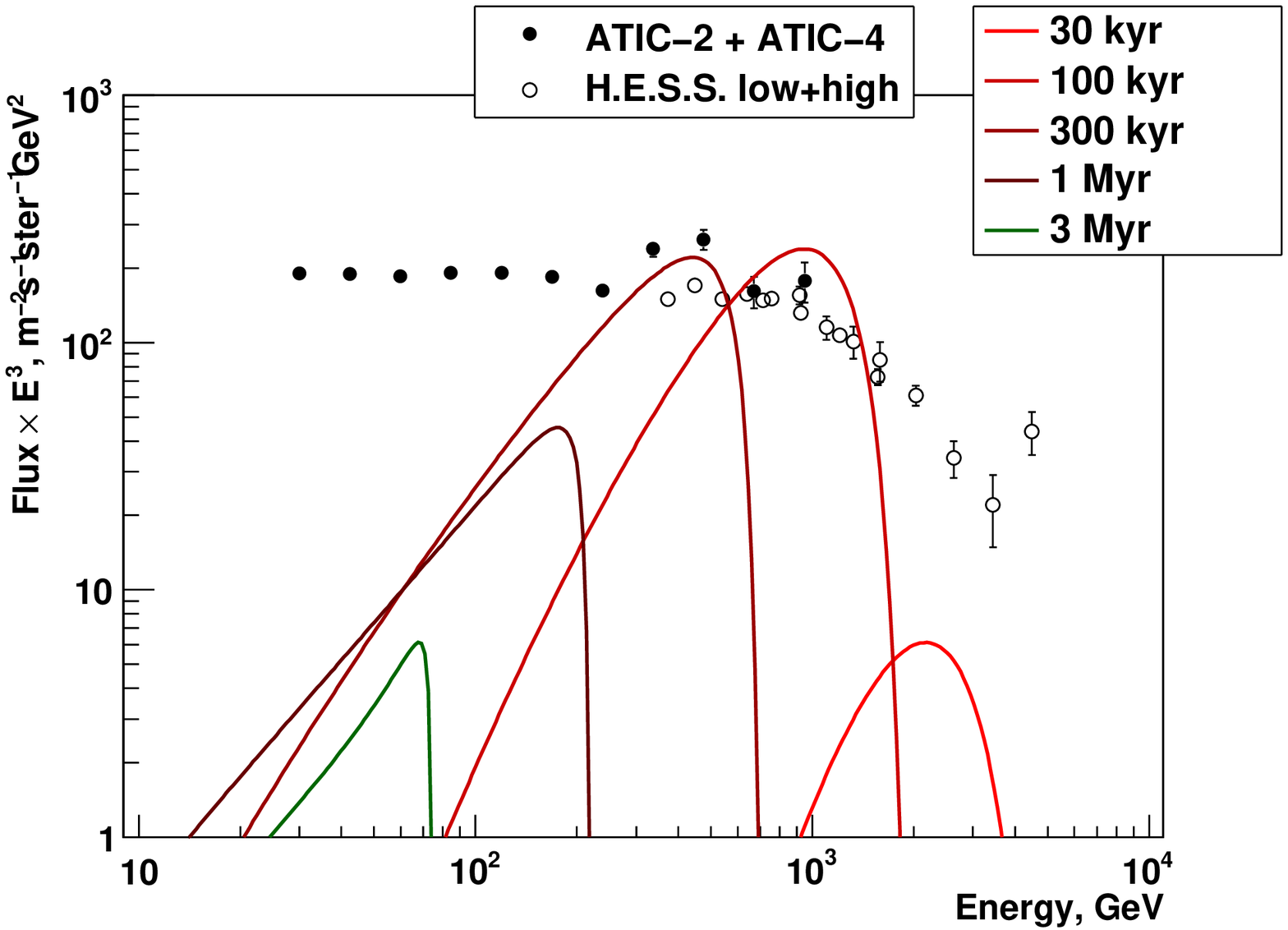}
\caption{\label{fig:Tr-T-R1000ECut01gamma1.3} 
A series of a single-pulsar electron spectra for different pulsar ages with the following other parametrs:
$r=1$\,kpc,
$\gamma=1.3$,
$E_{\mathrm{cut}}=1$\,TeV,
$\eta W_0=5\cdot10^{49}$\,erg,
$b_0=1.4\cdot10^{-16}$\,$(\mathrm{GeV}\cdot\mathrm{s})^{-1}$,
$D_0=3\cdot10^{28}$\,$\mathrm{cm}^2\cdot\mathrm{s}^{-1}$,
$\delta=0.3$.
The data of ATIC-2$+$ATIC-4 and H.E.S.S. (see \fig{AllElectronsCurrentState}) shown, also.
}
\end{minipage}\hfill
\end{figure}

The authors of paper \cite{CRE-THEOR-DELAHAYE2009} solved analytically the diffusion equation with suppositions which, generally, were quite similar to the ones of the conventional model  \cite{CRE-THEOR-PROTHEROE1982,CRE-THEOR-SM1998A}.
However, the efforts concentrated on the study of the origins of the theoretical uncertainties in the estimation of the positron flux (diffusion coefficient; cross sections; and other ISM parameters).
Therefore, the authors of \cite{CRE-THEOR-DELAHAYE2009} obtained not only one result but, also, a number of rather wide corridors of predictions.
In order to calculate the positron fraction, they used an experimental estimation for the total electron spectrum based mainly on the data of AMS-01 \cite{CRE-EXP-AMS01-2000}.
\fig{PositronsFraction-DelahayShaviv} shows the corridor of predictions of \cite{CRE-THEOR-DELAHAYE2009} in the supposition of $\gamma=3.35$.
It can be seen that the prediction cannot be reconciled with the data even with all the model uncertainties.
It is especially true with the last Fermi/LAT data for the positron fraction. 
We would like to note that the spectral index of $\gamma=3.35$ for the electron spectrum, used in the calculations, looks overestimated as shown by the last data of ATIC \cite{ATIC-2008-CHANG-NATURE}, Fermi/LAT \cite{CRE-EXP-FERMILAT2010B}, and PAMELA \cite{CRE-EXP-PAMELA2010-IzvFIAN}.
With a more realistic index $\gamma\approx3.0$, the model of paper \cite{CRE-THEOR-DELAHAYE2009} would predict an even steeper function for the positron fraction and the contradiction between the model and the data would be even firmer.


In the paper \cite{CRE-THEOR-PIRAN2009-PRL}, it was argued that inhomogeneity of cosmic ray sources, due to the concentration of SNRs towards the galactic spiral arms, could explain the anomalous increase of the positron fraction above 10~GeV. 
The idea was as follows. If the observer was located between galactic arms, then the primary electron flux produced by SNRs, which are located mainly in the arms, was suppressed due to radiative cooling. This was in contrast to the secondary positrons, sources of which -- cosmic ray protons -- were distributed homogeneously in the Galaxy.
The prediction of the model is shown in \fig{PositronsFraction-DelahayShaviv}. 
It can be seen that, actually, the model fits the data of PAMELA below 100~GeV. However, unknown to the authors of the paper \cite{CRE-THEOR-PIRAN2009-PRL}, it does not fit the more recent data of the Fermi/LAT observatory.
We would like to note, also, that, actually, the Sun is not located between galactic arms.
Instead, the Sun is located within the Orion-Cygnus arm \cite{APH-VAZQUEZ2008}. 
Therefore, strictly speaking, the model is inapplicable to the Sun position; however, nevertheless, it points out that, generally, the sources large-scale inhomogeneity should be taken into account.


Consequently, the last PAMELA and Fermi/LAT positron data seems to rule out the models in which the positrons are treated as purely secondary particles.
Therefore, some unconventional source of positrons should be incorporated into the models.


\subsection{Local burst-like sources of primary electrons and positrons: pulsars and SNRs}

\label{PULSARS}


There is a long history concerning the idea that, in the role of local nearby sources, pulsars, at high energies, could determine essential features of the electron spectrum and the positron fraction.
Starting in 1970, with C.S. Shen seminal work \cite{CRE-THEOR-SHEN1970},  nearby pulsars, as a possible source of cosmic ray electrons and structures in the electron spectrum, were discussed many times even before ATIC and PAMELA results (see for example \cite{CRE-THEOR-BOURALIS1989,CRE-THEOR-ATOYAN1995-PhysRevD,CRE-THEOR-KOBAYASHI-2001-ASR}).
It is important that pulsars produce high-energy negative electrons and positrons in equal parts (see for a review of the relevant physics of pulsars and pulsar wind nebulae (PWN) Appendix~A of \cite{CRE-THEOR-MALYSHEV2009B}, and, also, the paper \cite{CRE-THEOR-BLASI2010} and references therein).
Therefore, pulsars are definitely the sources of primary positrons which could be relevant to explaining the PAMELA anomaly in positron fraction.


There are, also, ideas that usual SNRs can accelerate not only negative electrons but, also, positrons.
Suprathermal positrons, in SNRs, can be produced in beta-decay in the chains starting with radioactive nuclei ejected during a supernova explosion: ${}^{56}$Ni, ${}^{44}$Ti, ${}^{26}$Al and ${}^{22}$Na \cite{CRE-THEOR-RAMATY1979-Nature,CRE-THEOR-RAMATY1979,CRE-THEOR-RAMATY1990,CRA-ZIRAKASHVILI-2011-PhysRedD,CRA-ZIRAKASHVILI-2011-ICRC}.
A possible role of reverse shocks, in the production and acceleration of negative electrons and positrons, was emphasized in \cite{CRA-ZIRAKASHVILI-2011-PhysRedD,CRA-ZIRAKASHVILI-2011-ICRC} and it was demonstrated that even the primary positron spectrum could be harder than the primary spectrum of negative electrons in \cite{CRA-ZIRAKASHVILI-2011-ICRC}.
It may be relevant to the PAMELA anomaly. 
Production and acceleration of secondary positrons in SNRs is considered in \cite{CRE-THEOR-BLASI2009,CRE-THEOR-MERTSCH2011}.
Independently, as regards the  production and acceleration of positrons in SNRs, it is argued in the literature over a long time, that nearby SNRs could produce complicated structures in the high-energy part ($\sim$1~TeV) of the cosmic ray electron spectrum \cite{CRE-EXP-NISHIMURA1997,CRE-THEOR-EW2002,CRE-THEOR-EW2009}. These may be relevant to the ATIC and PPB-BETS bump. 


In PWNe and SNRs, the character time of electrons acceleration is no more than 10--30~kyr and the character size of a relevant source is a few parsecs.
At the same time, the age of the sources, relevant to our observations of electrons in cosmic rays with TeV-energies and less, was greater than 200~kyr and the space scale of the observations was of the order or even much greater than 100~pc.
Therefore the approximation of point-like and instantaneous source is quite reasonable to estimate the electron flux from pulsars or SNRs. 
This approximation was introduced already in the C.~S. Shen 1970 paper \cite{CRE-THEOR-SHEN1970}, and, now, it is known as a burst-like approximation.


In the burst-like approximation, the source function for a source with spectrum $Q(E)$ located at $\mathbf{r}=0$ and $t=0$ is:
\begin{equation}
Q(\mathbf{r},t,E) = Q(E)\delta^3(\mathbf{r})\delta(t).
\label{eq:SourceBurst}
\end{equation}
By substituting the source function \eq{SourceBurst} to the Green-function solution of the transport equation \eq{DiffusionGenSol}, one obtains easily the observed density of electrons at arbitrary point $\mathbf{r}$ and arbitrary time $t>0$ in a simple analytical form:
\begin{equation}
\rho(\mathbf{r},t,E) = Q(E^*)\left(\frac{E^*}{E}\right)^2
  \frac{e^{-\mathbf{r}^2/2\lambda^2(E,E^*)}}{(2\pi)^{3/2}\lambda^3(E,E^*)};\quad
  E^*(E,t)=\frac{E}{1-b_0tE}.
  \label{eq:PulsarSolution}
\end{equation}
Here, $E^*(E,t)$ is the energy of an electron cooled down to $E$ due to the radiative losses during the time $t$.
Please note that $\rho(\mathbf{r},t,E)  \equiv 0$ for $E > E_{\mathrm{max}}(t)=1/(b_0t)$, therefore, there exists an exact cutoff energy $E_{\mathrm{max}}(t)$ which depends on the age of the source $t$. 


The most natural way to understand the PAMELA anomaly and the ATIC bump is to consider nearby pulsars as sources of electrons since, surely, they provide large fluxes of positrons.
To present a source spectrum $Q(E)$ of a pulsar the exponentially-truncated power-law spectrum is used widely:
\begin{equation}
 Q(E) = Q_0 E^{-\gamma} \exp(-E/E_{\mathrm{cut}}).
 \label{eq:QPulsar}
\end{equation}
In the source function, there are three free papameters -- $Q_0, \gamma, E_{\mathrm{cut}}$  -- which should be selected on the basis of observations or from some other additional considerations.


The source spectral index $\gamma$ for a PWN, is estimated  to be $1.3\pm0.3$ on the  basis of radio astronomic data for seven pure PWNe  derived from the last version of D.A. Green catalogue \cite{APH-GREEN2009} (see, also, the discussion in \cite{CRE-THEOR-MALYSHEV2009B}).
It can be seen that PWN electron source spectra are harder than electron source spectra of SNRs ($\gamma_{\mathrm{SNR}}\gtrsim2.0$). 


For $\gamma < 2$, the amplitude $Q_0$ may be estimated from  the pulsar complete energy deposit to electrons using the following simple formula:
\begin{equation}
 \int_0^\infty Q(E) E dE = Q_0 \Gamma(2-\gamma) E_{\mathrm{cut}}^{2-\gamma} = \eta W_0,
\end{equation}
Where $W_0$ is the spin-down energy of the pulsar which transforms mainly to the energy of electromagnetic radiation and energy of electron-positron pairs, and $\eta$ is the conversion factor for spin-down energy to transform to the energy of electron-positron pairs.


There are significant uncertainties in the estimations of $W_0$ for known pulsars. However, generally, it is assumed that, for the majority of pulsars, $W_0$ is between $5\cdot10^{48}$ and $5\cdot10^{50}$~erg \cite{CRE-THEOR-MALYSHEV2009B}, \cite[Table 4]{CRE-THEOR-DELAHAYE2010}.
The conversion coefficient $\eta$ is expected to be $\sim1$ but, effectively, it could be somewhat less (down to $\sim0.1$) due to energy losses during the retention of electrons within PWN before destruction of the PWN and ejection of the electrons into the ISM \cite{CRE-THEOR-MALYSHEV2009B}.
Therefore $10^{48}$--$10^{50}$~erg is a reasonable estimate for $\eta W_0$.


The cutoff energies $E_{\mathrm{cut}}$ are not known well. However, there are indications that they might be as high as several tens of TeV.
For example, the electron cutoff energy for Vela PWN, as estimated by the gamma-telescope H.E.S.S., is 67~TeV \cite{CRGAMMA-HESS-2006-AA}\footnote{However, the estimated spectral index from gamma-ray flux is $\gamma\approx2.0$ and Vela is not pure PWN according to the catalogue \cite{APH-GREEN2009}, it is PWN$+$SNR nebula.}.


\begin{figure}
\begin{center}
\includegraphics[width=\pictsize]{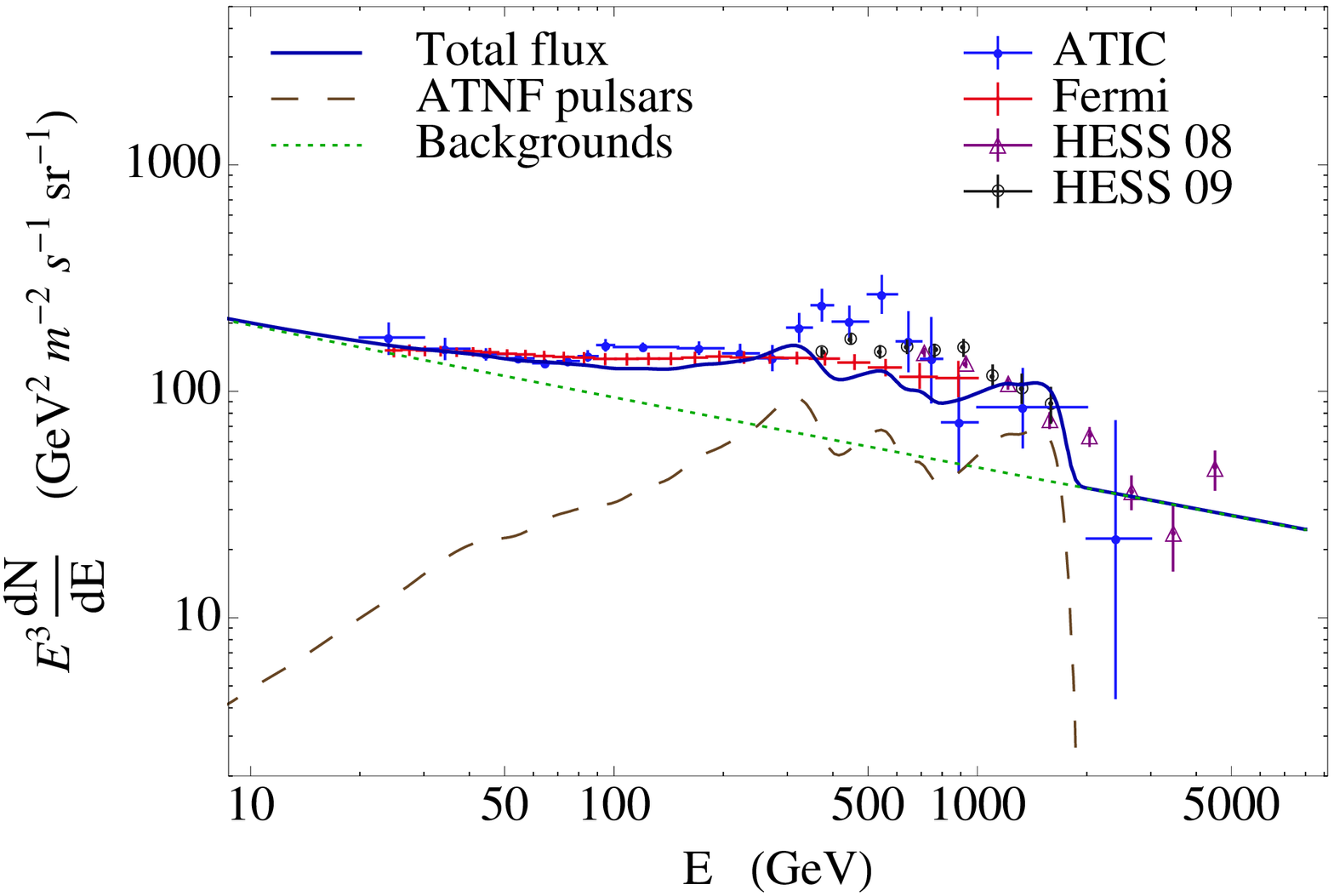}
\includegraphics[width=\pictsize]{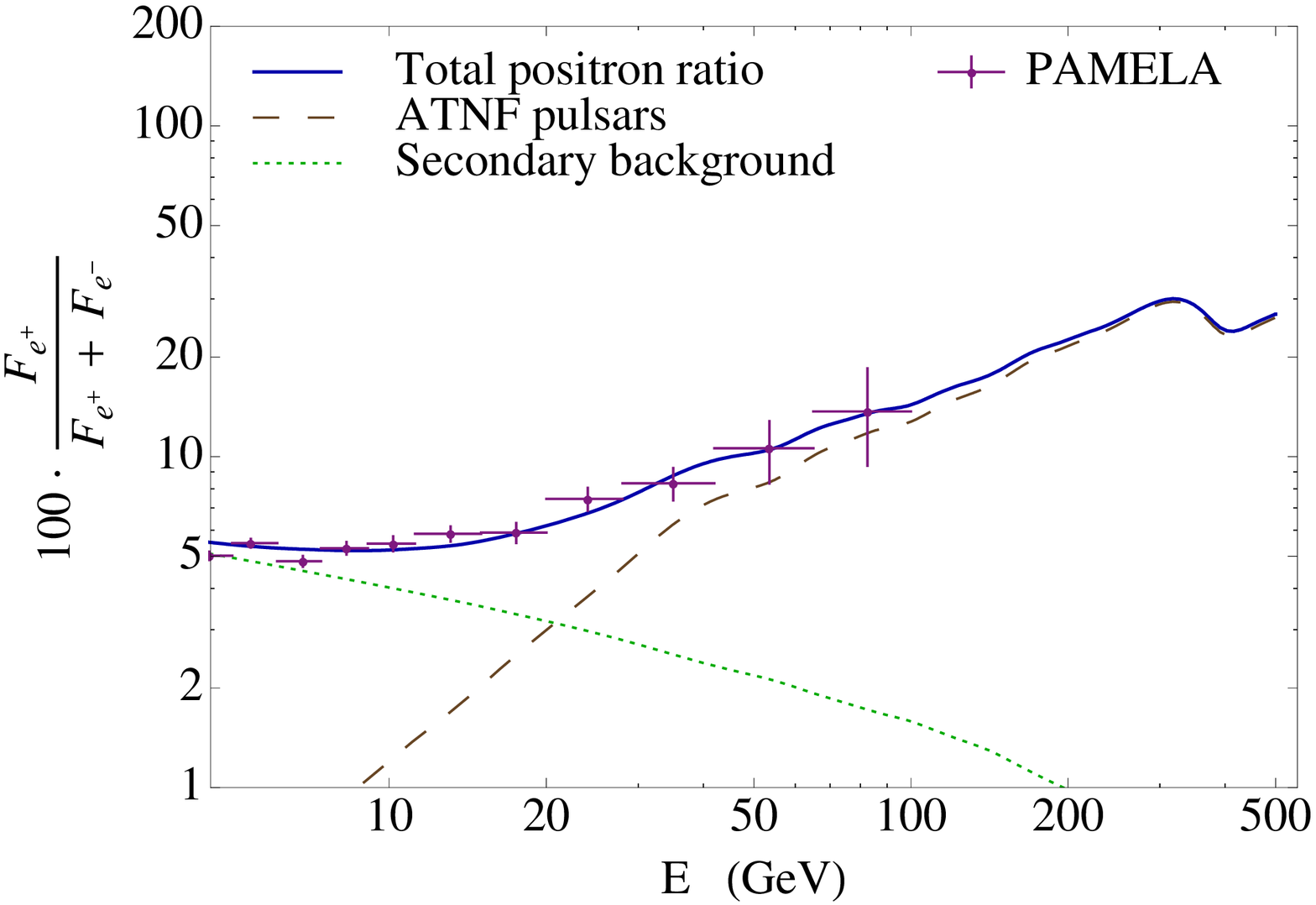}
\end{center}
\caption{\label{fig:Malyshev-Pulsars} 
The predicted electron spectrum and positron fraction for pulsars in the ATNF catalog together with conventional background as calculated in \cite[FIG.4]{CRE-THEOR-MALYSHEV2009B} (the figure reproduced by permission of the authors).
}
\end{figure}


By variation of the parameters $r,t,Q_0,\gamma,E_{\mathrm{cut}}$, one could obtain a variety of different single-pulsar spectra with various shapes and intensities.
\fig{Tr-T-R1000ECut01gamma1.3} shows an example of a series of a single-pulsar electron spectra with different ages but with the same other parameters.
It can be understood from \fig{Tr-T-R1000ECut01gamma1.3} that within the existing uncertainties of parameters of known nearby pulsars the intensity and shape of the spectrum provided by a combination of several pulsars together with conventional background can be agreed with the data of ATIC, Fermi/LAT, H.E.S.S. etc. by a variety of ways. 
Really, many recent papers describe the data by this manner (see for example \cite{CRE-THEOR-PROFUMO2008,CRE-THEOR-GRASSO2009,CRE-THEOR-MALYSHEV2009B,CRE-THEOR-BLASI2010,CRE-THEOR-DELAHAYE2010,CRE-THEOR-PROFUMO2010}; for review see \cite{CRE-THEOR-YiZhongFan2010}).


In order to describe the electron spectrum high energy features in such models, an essential part or, even, the most of the flux, at energies $E \gtrsim 200$~GeV, is given by a contribution of nearby pulsars with equal parts of positrons and negative electrons.
Therefore, generally, the positron fraction increases from the low values like $\sim0.05$ at $E\approx10$~GeV up to the values $\gtrsim0.2$ at energies 300--500~GeV.
Consequently, such models explain simultaneously the PAMELA positron anomaly together, very naturally, with the no-power-law behaviour  of the electron spectrum.
This result was demonstrated in all cited above papers \cite{CRE-THEOR-PROFUMO2008,CRE-THEOR-GRASSO2009,CRE-THEOR-MALYSHEV2009B,CRE-THEOR-BLASI2010,CRE-THEOR-DELAHAYE2010,CRE-THEOR-PROFUMO2010} and, also, in the earlier works \cite{CRE-THEOR-AHARONIAN1995-A+A,CRE-THEOR-ATOYAN1995-PhysRevD}. 
The calculations for all pulsars in the ATNF catalogue carried out in the Malyshev et al. paper \cite{CRE-THEOR-MALYSHEV2009B}, are shown as an example in \fig{Malyshev-Pulsars}.


The possibility of SNRs to accelerate not only electrons but, also, positrons does not interfere with working models of pulsars as sources of electrons.
Moreover, for the same reasons as pulsar models, the acceleration of positrons only by SNRs, without pulsars at all, can explain, in principle, the electron spectrum and the positron fraction simultaneously \cite{CRE-THEOR-BLASI2009}.
In all these cases, the models kernel part is the possibility of local instantaneous sources to accelerate both negative electrons and positrons.


\subsection{Dark matter annihilation and decay}


WIMPs annihilation (or decay) is the third main way in understanding the features of the electron spectrum and the positron fraction anomaly.
WIMP means `weakly interacting massive particle' --  one of the most promising candidates for the DM (see for review \cite{CREDM-GRIEST1995,CREDM-JUNGMAN1996,CREDM-ARKANI2008,CREDM-HOOPER2009}). 
For the first time the possible anomaly in the positron fraction at energies above 10~GeV was directly attributed to the annihilation of WIMPs in the paper of A. J. Tylka \cite{CREDM-TYLKA1989} of 1989. 
Later, even before 2000, this question was studied several times \cite{CREDM-TURNER1990,CREDM-TURNER1991,CREDM-BALTZ1999,CRE-THEOR-COUTU1999} and hundreds of papers were published  following ATIC and PAMELA results.
In the electron spectrum, the bump-like feature was considered to be a possible signature of WIMPs annihilation to the electron-positron pairs already in the PPB-BETS paper \cite{CRE-EXP-PPB-BETS2008}. However, it was noticed that the statistics were too low to make definite conclusions.
For the first time in the electron spectrum, the observed bump was attributed clearly directly to a possible signature of the annihilation of DM particles in the ATIC paper \cite{ATIC-2008-CHANG-NATURE}.The mass of the particle was estimated as $\approx620$~GeV\footnote{It was done in \cite{ATIC-2008-CHANG-NATURE} for the case of Kaluza-Klein particles which produce one electron-positron pair per annihilation.}.
Now there are numerous papers which consider the no-power-like shape of the electron spectrum by similar ways (see for review \cite{CRE-THEOR-YiZhongFan2010})


Many-body simulations show that WIMP-like DM halo of the Galaxy must be clumpy \cite{CREDM-DIEMAND2008-Nature}.
In the paper \cite{CREDM-OLINTO2001}, it was noted, for the first time, that the DM clumps (subhalos) could enhance many times the DM annihilation rate, and, therefore, the clumpы of the DM was important for the problem of cosmic ray electrons ejection.
Consequently, there are two different but connected problems: firstly, the observation of electrons produced by the smooth Galactic DM halo and, secondly, by nearby local clumps.
We considered two distinct analytical solutions of the electron transport equation \eq{DiffusionEq} related to these problems. 


\subsubsection{Smooth Galactic DM halo.}
\label{SMOOTH-HALO}


In order to obtain a simple analytical solution related to a smooth Galactic halo, we supposed the source of electrons to be infinitely large in space and homogeneous, with the source spectrum $Q(E)$:
\begin{equation}
 Q(\mathbf{r},t,E) = Q(E).
 \label{eq:SourceHalo}
\end{equation}
By substituting the source function \eq{SourceHalo} to the general Green function solution \eq{DiffusionGenSol} of the transport equation, one obtains easily the solution for the observable electron spectrum:
\begin{equation}
 \rho(E) = \frac{1}{b_0E^2}\int_E^\infty Q(E')dE'.
 \label{eq:SourceHaloGenSol}
\end{equation}
The energy spectrum of electrons, ejected in the annihilation of WIMPs, could be delta-like in the simplest case or, in general, more complicated. We considered only the simplest delta-like case.
The results for more general situations could be understood easily on the basis of the simplest one. The source spectrum for delta-like annihilation spectrum is
\begin{equation}
 Q(E) = Q_0\delta(E-E_0),
 \label{eq:SourceHaloDelta}
\end{equation}
and from \eq{SourceHaloGenSol} one immediately obtains
\begin{equation}
  \rho(E) =\left\{
  \begin{array}{rl}
  \displaystyle \frac{Q_0}{b_0}\frac{1}{E^2},&\quad E<E_0\\
  0, &\quad E>E_0
  \end{array}
  \right.
  \label{eq:SolutionHaloDelta}
\end{equation}

\begin{figure}
\begin{minipage}[t]{\htw}
\includegraphics[width=\pictsize]{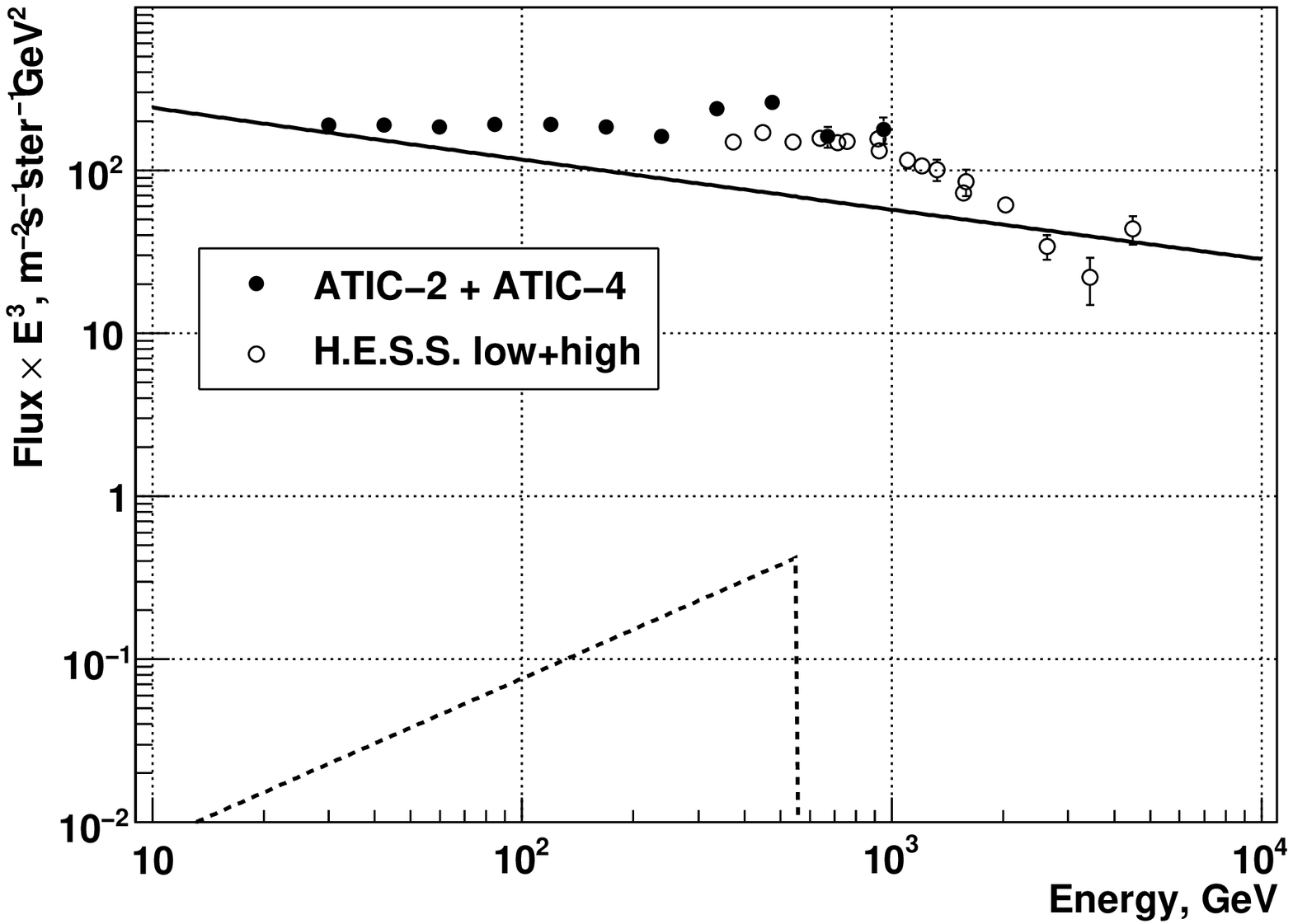}
\end{minipage}%
\hfill
\begin{minipage}[b]{\htw}
\caption{\label{fig:DM-Halo-Single} 
A typical prediction of a simple analytical model \eq{SolutionHaloDelta} for the electron spectrum produced by homogeneous DM-halo without boost factor.
Dashed line -- the spectrum of electrons from DM annihilation, the solid line -- the sum of DM spectrum and conventional background.
Dark matter peak is invisible above the background.
The parameters of the model are: $M_{\mathrm{WIMP}}=1.1$~TeV, $k=4$, $\langle\sigma v\rangle=3\cdot10^{-26}\mathrm{cm}^3\mathrm{s}^{-1}$, $n=0.3\,\mathrm{GeV}\cdot\mathrm{cm}^{-3}/M_{\mathrm{WIMP}}$, $Q_0=4.5\cdot10^{-33}\,\mathrm{cm}^{-3}\mathrm{s}^{-1}$.
}
\end{minipage}\\
\begin{minipage}[t]{\htw}
\includegraphics[width=\pictsize]{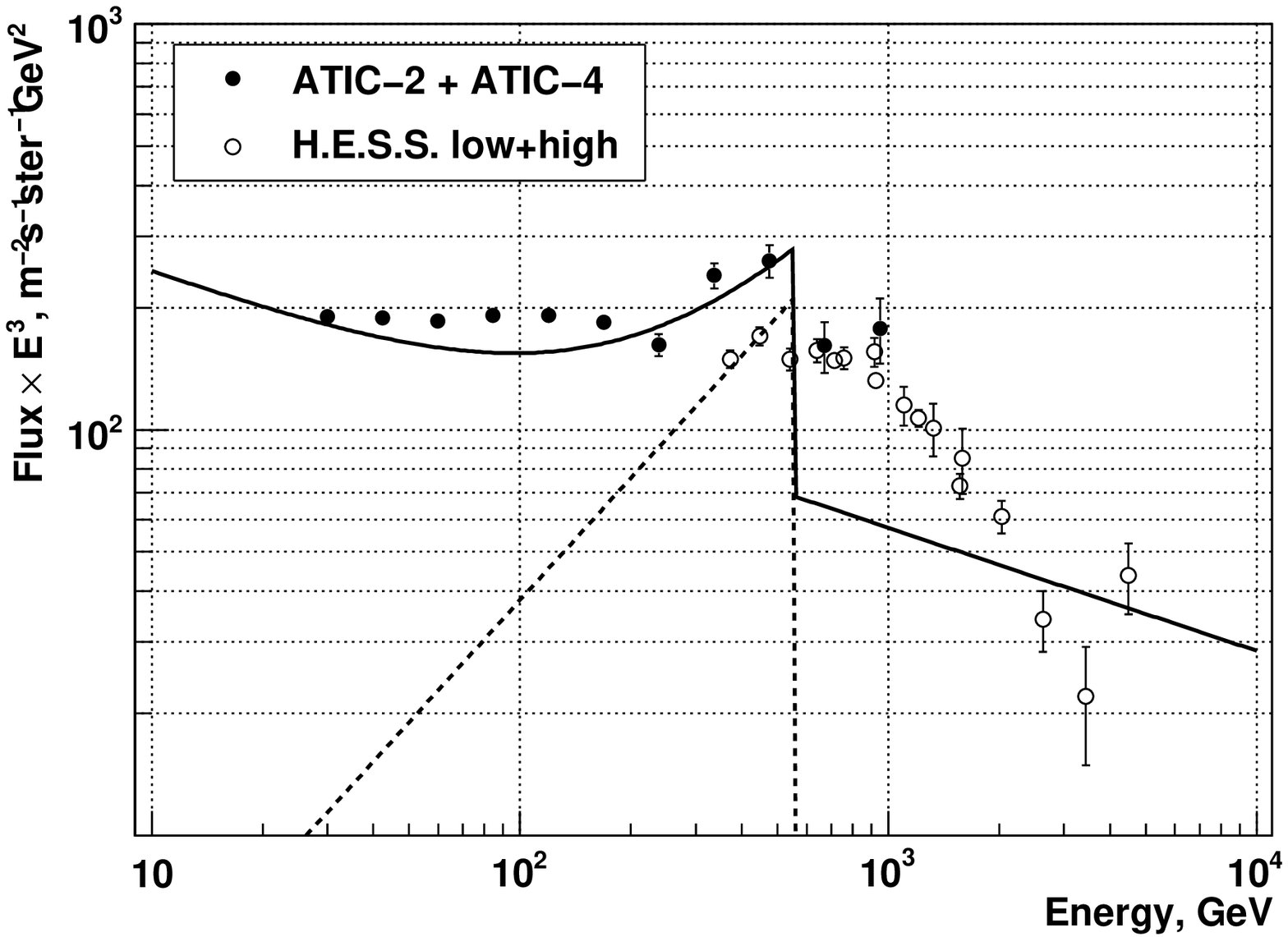}
\caption{\label{fig:DM-Halo-Single-Boost300} 
Same as \fig{DM-Halo-Single} but with boost factor 500.
}
\end{minipage}\hfill
\begin{minipage}[t]{\htw}
\includegraphics[width=\pictsize]{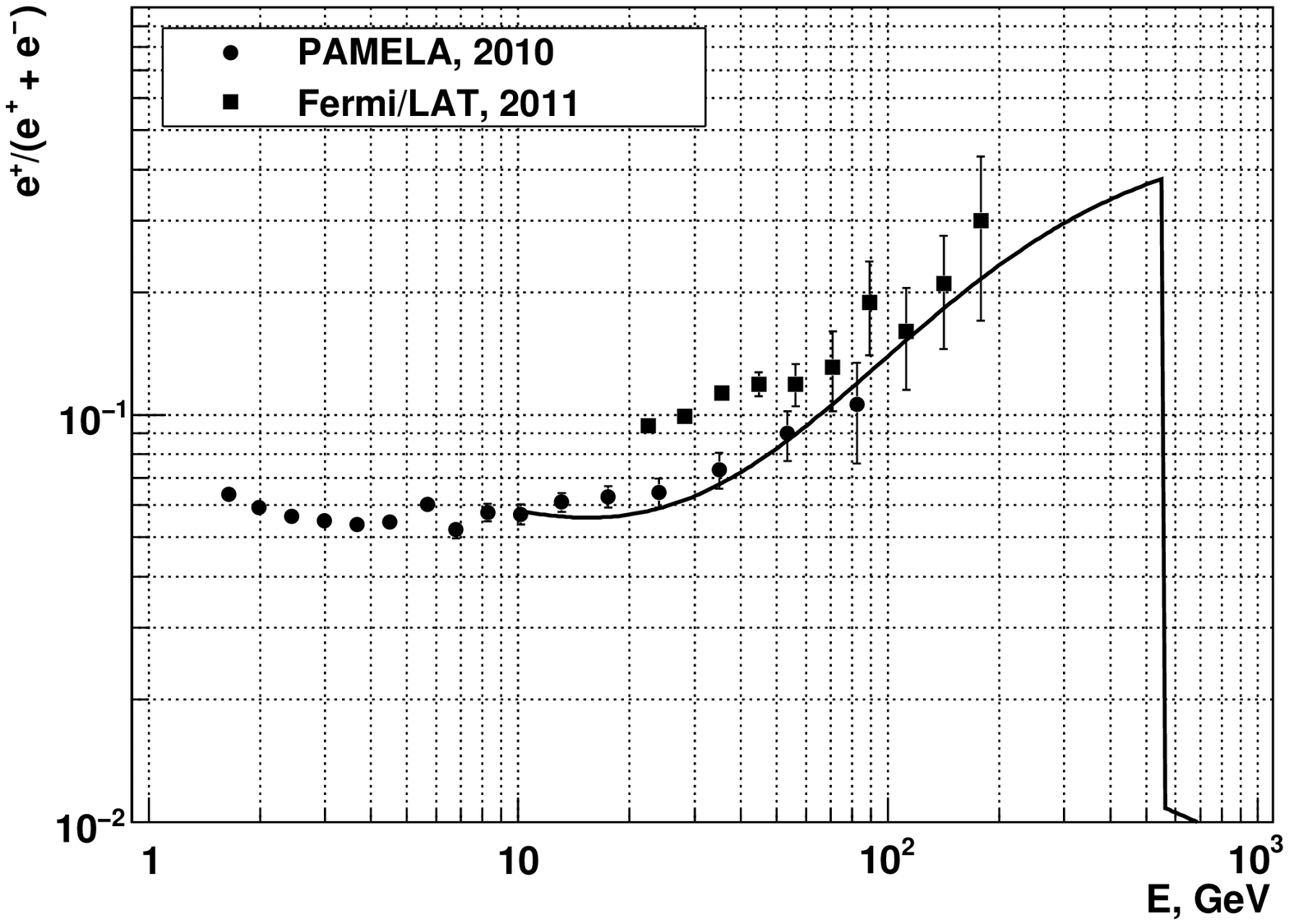}
\caption{\label{fig:DM-Halo-Single-Boost300-positrons} 
The predicted positron fraction for the \fig{DM-Halo-Single-Boost300} model.
}
\end{minipage}\hfill
\end{figure}

The rate of annihilation $Q_0$ in the formulae \eq{SourceHaloDelta} and \eq{SolutionHaloDelta}  may be estimated as
\begin{equation}
 Q_0 = \frac{1}{2}kn^2\langle\sigma v\rangle,
 \label{eq:HaloDeltaQ0}
\end{equation}
where $k$ is the number of electrons per one annihilation; $n\,[\mathrm{cm}^{-3}]$ is the WIMPs density in the smooth Galactic halo; $\sigma$ is the cross section of annihilation; and $v$ is relative velocity of WIMPs. 
Here, we assume, also, for simplicity that the WIMP particle is its own antiparticle (like Majorana fermions).
The density $n$ is estimated as $n\sim0.3\,\mathrm{GeV}\cdot\mathrm{cm}^{-3}/M_{\mathrm{WIMP}}$; and, for the WIMPs annihilation cross section, it is known from the cosmological abundance of DM $\langle\sigma v\rangle\sim3\cdot10^{-26}\mathrm{cm}^3\mathrm{s}^{-1}$ \cite{CREDM-JUNGMAN1996}. This value is expected to be independent on $v$ for annihilation in $S$-mode (the Bethe law).


As an example, we consider here the annihilation of WIMPs to two electron-positron pairs and suppose $M_{\mathrm{WIMP}}=1.1$~TeV. 
Then $k=4$ and $n=2.7\cdot10^{-4}\,\mathrm{cm}^{-3}$. 
From \eq{HaloDeltaQ0} one obtains easily $Q_0=4.5\cdot10^{-33}\,\mathrm{cm}^{-3}\mathrm{s}^{-1}$ and the respected spectrum calculated with equation \eq{SolutionHaloDelta} is shown in \fig{DM-Halo-Single}. 


It can be seen that the amplitude of the DM spectrum is almost three orders of magnitude less than the amplitude of experimental electron spectrum and that the DM deposit is even invisible above the conventional background.
This is due to too low cross section value $\langle\sigma v\rangle\sim3\cdot10^{-26}\mathrm{cm}^3\mathrm{s}^{-1}$ defined by the cosmological data and it is generic feature of all DM models of an origin of cosmic ray electrons. 
In order to obtain reasonable amplitude for the DM annihilation deposit to the electron spectrum, one should suppose some mechanism of increasing the annihilation cross section $\langle\sigma v\rangle$ at low energies -- the so-called cross section boost factor.
Possible mechanisms may be either a Sommerfeld effect or a Breit-Wigner enhancement factor (see for review \cite{CREDM-XIAOGANG2009A}).
\fig{DM-Halo-Single-Boost300} shows the DM spectrum with the same parameters, as described above (see also \fig{DM-Halo-Single}) but with a 500 boost factor.
In \fig{DM-Halo-Single-Boost300}, the very sharp peak is not a problem of this model but only an artifact of a delta-like annihilation spectrum of electrons \eq{SourceHaloDelta}.
For a more smooth annihilation spectrum, the observed spectrum would be a simple convolution of the peaks like in \fig{DM-Halo-Single-Boost300} with the actual source annihilation spectrum and the shape of the predicted spectrum being smoother.


Negative electrons and positrons are produced in the annihilation of WIMPs in equal parts. 
Therefore, if some DM model predicts no-power-law spectrum or bumps in high-energy part of the electron spectrum with sufficiently high amplitude such as in \fig{DM-Halo-Single-Boost300}, it produces almost automatically reasonable behaviour of the positron fraction, as shown in \fig{DM-Halo-Single-Boost300-positrons}.
This important feature of the DM models does not depend on details like the annihilation spectrum etc., and the feature is very general.
The origin of this feature is similar to the origin of the similar feature of the pulsars models (see section \ref{PULSARS} and \fig{Malyshev-Pulsars}).
Many papers (\cite{CREDM-CIRELLI2008,CREDM-CHEN2008,CREDM-XIAOGANG2009A} and others, for review see \cite{CRE-THEOR-YiZhongFan2010}) demonstrated the possibility of smooth halo DM models to predict, at the same time, bumps in the electron spectrum and the reasonable positron fraction consistent with the PAMELA anomaly.


\subsubsection{Subhalos (clumps)}
\label{SUBHALOS}


From the late 1990's, it  was known from numerical many-body simulations  \cite{CREDM-GHIGHA1998,CREDM-KLYPIN1999} that the Galactic dark mater halo had to be clumpy since there was an incomplete merging of primordial DM substructures which formed the Galactic halo.
Recent many-body simulations, with the code Via Lactea II \cite{CREDM-DIEMAND2008-Nature}, showed that the subhalos (clumps) were distributed with approximately equal total mass per decade of subhalo masses over the range $10^6M_\odot$--$10^9M_\odot$, where $M_\odot$ was the Solar mass.
Within the sphere with a 400~kpc radius with the center at the core of the  Milky Way, it is expected that there is $\sim10$ subhalos of $\sim10^9M_\odot$, $\sim10^4$ subhalos of $\sim10^6M_\odot$ etc.
The mean distance, between $10^6M_\odot$ subhalos, is $\sim10$~kpc and, therefore, it is approximately the expected distance from the Sun to the nearest subhalo of this mass.
The mean DM  density, within such a clump, is expected to be hundreds of times greater than the mean density of the smooth Galactic halo, and the density near the central parts of a clump (a cusp) -- is expected to be thousands of times greater.
The annihilation rate is proportional to the square of the DM density. Therefore, there is a natural `boost factor' for the annihilation rate and the related flux of electrons may be rather high near the Sun. This problem should be investigated carefully.


\begin{figure}
\begin{minipage}[t]{\htw}
\includegraphics[width=\pictsize]{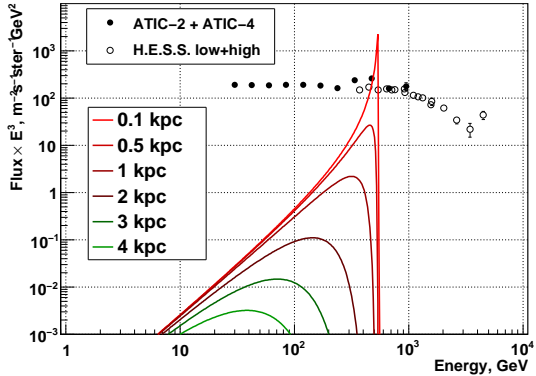}
\end{minipage}%
\hfill
\begin{minipage}[b]{\htw}
\caption{\label{fig:DM-Clump-RTrace} 
A family of electron spectra produced by a single DM clump without boost factor for different distances from an observer to the clump. The total mass of the clump is about $2.5\cdot10^6M_\odot$, the total annihilation rate is $Q_0 = 1.7\cdot10^{34}\mathrm{s}^{-1}$ with delta-like spectrum of electrons concentrated at $550$~GeV.
}
\end{minipage}
\end{figure}


The scale of sizes of $\sim10^6M_\odot$ clumps is $\sim100$~pc, the scale of sizes of most dense and bright cusps of clumps is only $\sim10$~pc \cite{CREDM-DIEMAND2008-Nature}.
A point-like approximation is quite reasonable for the observations of the clump from distances $\sim1$~kpc and greater.
Also, in time, the related source of electrons should be considered as permanent and constant.
Therefore, supposing as above a delta-like spectrum for annihilation electrons, we consider the source function to be: 
\begin{equation}
 Q(\mathbf{r},t,E) = Q_0\delta^3(\mathbf{r})\delta(E-E_0).
 \label{eq:SourceClump}
\end{equation}
Substituting the source function \eq{SourceClump} to the Green function solution \eq{DiffusionGenSol} of the transport equation one obtains easily a simple analytical solution which helps to understand the DM clumps physics related to cosmic ray electrons:
\begin{equation}
 \rho(\mathbf{r},t) =
 Q_0\, \frac{\exp[-r^2/2\lambda^2(e,E_0)]}{b_0 E^2 (2\pi)^{3/2}\lambda^3(E,E_0)}.
\end{equation}

In order to estimate an observed electron spectrum from a dark-matter clump, we supposed a simple clump toy model.
The clump was represented by two concentric spheres.
The outer sphere presented a volume of the clump; the inner sphere presented a cusp.
For the volume, we supposed $\rho_{DM}(\mathrm{volume}) = 2.5\,M_\odot/\mathrm{pc}^3$, $V(\mathrm{volume}) = (100\mathrm{pc})^3$; 
for the cusp we supposed $\rho_{DM}(\mathrm{cusp}) = 45\,M_\odot/\mathrm{pc}^3$, $V(\mathrm{cusp}) = (10\mathrm{pc})^3$. 
The total mass of such clump was about $2.5\cdot10^6M_\odot$. 
These parameters looked possible for a clump with such sizes (see \cite{CREDM-DIEMAND2008-Nature}); however, the density and mass could be, also, a $\sim5$ times less \cite{CREDM-DIEMAND2008-Nature}).
Therefore, our toy model clump was relatively dense and bright and our estimates for the amplitude of the electron spectrum should be considered as very optimistic.


For this toy model, with the parameters of DM particles as in the Section \ref{SMOOTH-HALO}, one obtains easily $Q_0 = 1.7\cdot10^{34}\mathrm{s}^{-1}$ in the eq.~\eq{SourceClump}.
\fig{DM-Clump-RTrace} shows a family of the spectra for various distances from the Sun to the clump.
It can be seen that, even without  a boost factor, the deposit from the nearby DM clump to the electron spectrum may be observable, in principle, but only for a very close clump $r \lesssim 0.1$\,kpc (in fact -- within the clump).
This situation is very unrealistic. For more realistic distances $r\sim1\div10$~kpc the clump deposit is negligibly small compared to the observable flux of electrons and it may be observed only in the case of high boost factors like $10^2\div10^3$. 


If, actually, the annihilation energy spectrum of the electrons is delta-like, then, it is possible, in principle, to obtain a very sharp peak in the electron spectrum.
This situation may take place in the case of a DM clump located very close to the observer (see \fig{DM-Clump-RTrace}).
However, we would like to emphasize that it is impossible, in principle, to have two or more sharp peaks related to DM at different energies and at the same time.
As clear from \fig{DM-Clump-RTrace}, if there are two or more clumps at different distances from the Sun, they must produce peaks at different energies (due to cooling of the electrons). However, only the nearest clump could produce a sharp peak, the others produce only very smooth distribution.
It was noted in \cite{CRE-THEOR-MALYSHEV2009B} and direct calculations, for many-clump structures, revealed, also, this feature \cite{CREDM-BRUN2009,CREDM-BRUN2009-PhysRev}.


\begin{figure}
\begin{center}
\includegraphics[width=\pictsize]{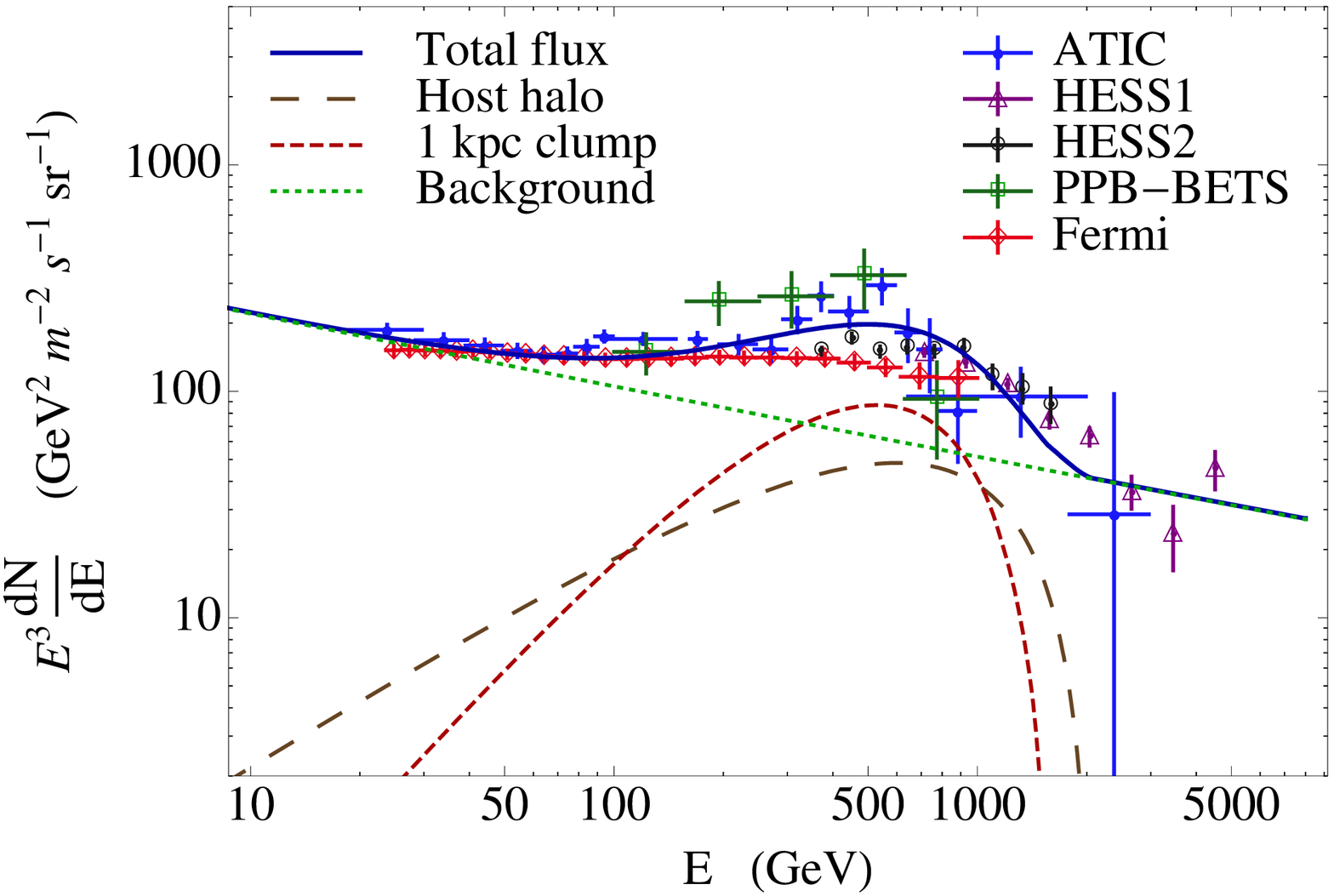}
\includegraphics[width=\pictsize]{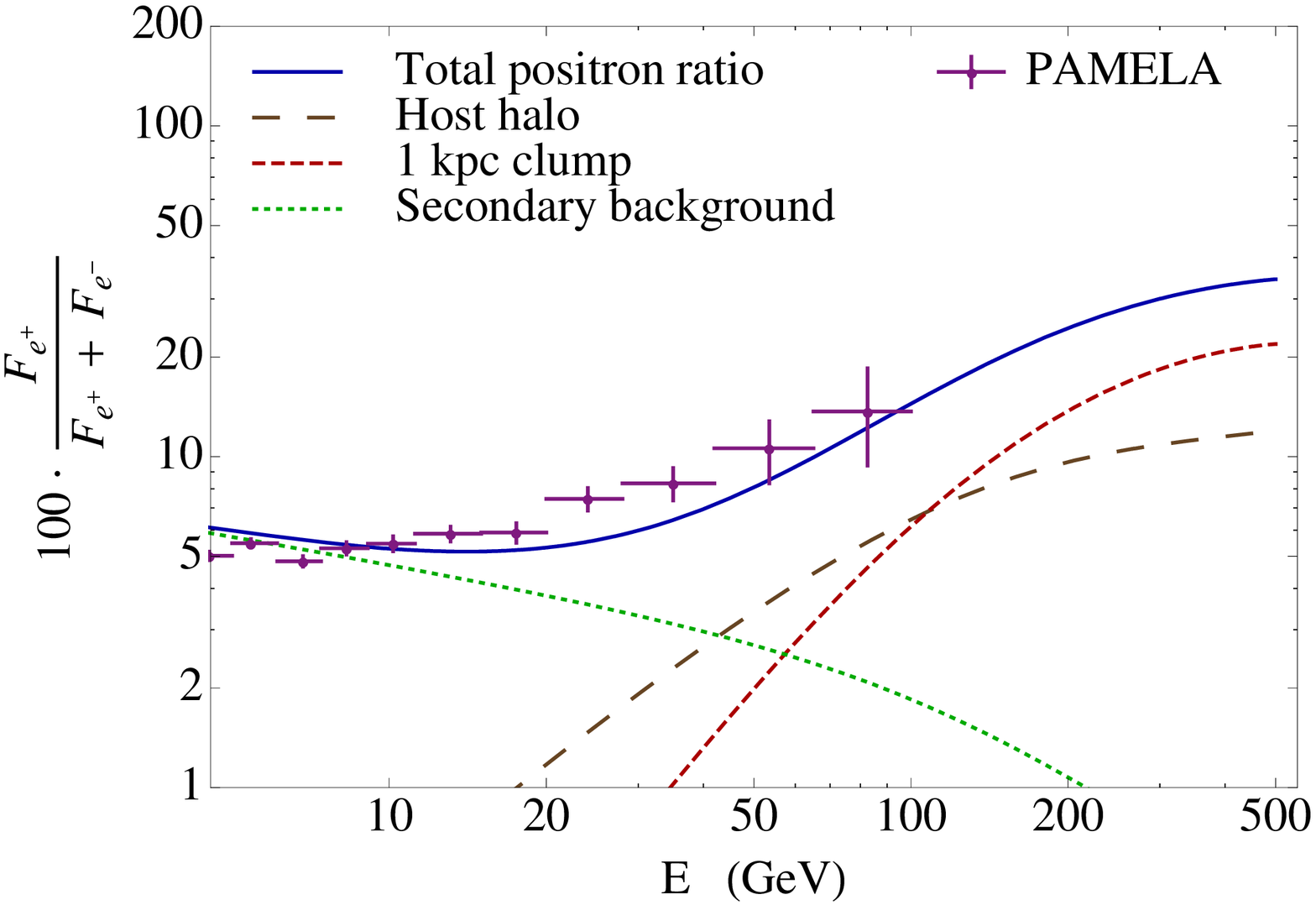}
\end{center}
\caption{\label{fig:Malyshev-DM}
Toatal electron flux and positron fraction from boosted annihilating DM in the host halo plus 1~kpc-clump as calculated in \cite[FIG.7]{CRE-THEOR-MALYSHEV2009A} (the figure reproduced by permission of the authors). See details of the calculation in \cite{CRE-THEOR-MALYSHEV2009A}.
}
\end{figure}


A possibility to predict simultaneously bump-like structures in the electron spectrum and rising positron fraction, is a generic property of subhalo DM models.
This is an implication of the equal fluxes of negative electrons and positrons in the annihilation of DM particles and it is quite similar to the same possibility of the pulsars models and DM smooth halo models (see above).
However, the spectrum, produced by a DM clump, is generally harder than the spectrum produced by a smooth halo.
The result is a deficit of positrons near the energies $\sim100$~GeV relative to PAMELA data for a pure DM clump scenario \cite{CRE-THEOR-MALYSHEV2009A}. 
However, the dark matter smooth halo, together with a deposit of clumps, can describe bumps in the electron spectrum and, at the same time, the anomaly in the positron fraction.
It was demonstrated in many papers (see for example \cite{CREDM-BRUN2009,CREDM-BRUN2009-PhysRev,CRE-THEOR-MALYSHEV2009A}, for review, please see \cite{CRE-THEOR-YiZhongFan2010}).
As an example of such predictions, \fig{Malyshev-DM} shows the total flux of the host smooth halo, together with a nearby 1~kpc clump calculated in the paper \cite{CRE-THEOR-MALYSHEV2009A}.


We did not discuss in detail the possibility of explaining the ATIC and PAMELA anomaly by decay of DM particles with generation of electron-positron pairs.
Such a possibility is discussed widely in the literature (see for example \cite{CREDM-BAE2009} and references therein).
This explanation shares many features of DM annihilation with the exception of a strong boosting due to a high concentration of DM in local clumps.
Particularly, DM decay cannot produce several distinct sharp peaks in the electron spectrum as well as DM annihilation cannot, as was discussed above\footnote{In order to produce several distinct sharp peaks in the spectrum of electrons, the only possibility is to have, at the same time, 1) DM decay or annihilation with several distinct delta-like peaks in the energy spectrum of electrons; and 2) location of the Sun inside a DM-clump, close to its cusp. It looks absolutely improbable.}.


\subsection{To the model selection}


It can be seen from the review, given in Section~\ref{THREE-WAYS} that the conservative way could not explain the positron anomaly; therefore, it was ruled out.
On the contrary, both the pulsars scenario and the DM scenario could explain successfully the data of low-resolution experiments (see definition of low-resolution experiments in Section 3.1).
Therefore, it was impossible to select between these two explanations only on the  basis of low-resolution electron data. 


In the DM scenario for electrons, there are several troubles.
A natural prediction of DM models is an anomaly in the antiproton fraction in cosmic rays similar to the positron anomaly; however, a similar anomaly was not observed in the PAMELA experiment \cite{CRMAGNET-PAMELA-2009-AntiP,CRMAGNET-PAMELA-2010-AntiP}.
This trouble is avoidable by suggestion of a leptophilic annihilation of DM particles \cite{CREDM-YIN2008}.
The strongest constrains on the DM annihilation models come from gamma-ray astronomy. 
There are large uncertainties associated with model-dependent or poorly known astrophysical factors like DM density profiles. Nevertheless, the obtained upper limit to $\langle\sigma v\rangle$ excludes the smooth Galactic halo as the only source of the ATIC and PAMELA anomaly \cite{CREDM-GAMMA-ZAVALA-2011,CREDM-GAMMA-HESS-2011}.
However local DM substructures like clumps are not excluded by gamma-ray constraints since they may in principle produce strong signal even without boost factor (see \fig{DM-Clump-RTrace}). 

However, local DM substructures, like clumps, are not excluded by gamma-ray constraints since, in principle, they may produce strong signals even without a boost factor (please see \fig{DM-Clump-RTrace}).
Therefore, DM models cannot be excluded and the nature of electron anomalies remains unclear.


The situation may be changed dramatically if the results of high-resolution ATIC experiment \cite{ATIC-2011-PANOV-ASTRA}, considered in the next section, are confirmed. 


\section{The fine structure of the electron spectrum measured by the ATIC spectrometer}
\label{FINESTRUCTURE}


The ATIC (Advanced Thin Ionization Calorimeter) balloon-borne spectrometer was designed to measure the energy spectra of nuclei from hydrogen to iron with an individual resolution of charges in primary cosmic rays for energy region from 50 GeV to 100 TeV. 
It was shown that ATIC was capable of not only measuring the spectra of cosmic ray nuclear components but, also, the spectrum of cosmic ray electrons \cite{ATIC-2008-CHANG-ADS}.
In order to separate the electrons from a much higher background of protons and other nuclei, the differences in the shower development in the apparatus for electrons and nuclei are used. 
The low-resolution spectrum (see the definition of the term in Section~\ref{ELECRTON-SPECTRUM}), measured by ATIC in this way, was published in the paper \cite{ATIC-2008-CHANG-NATURE}.


\begin{figure}
\begin{minipage}[t]{\htw}
\includegraphics[width=\pictsize]{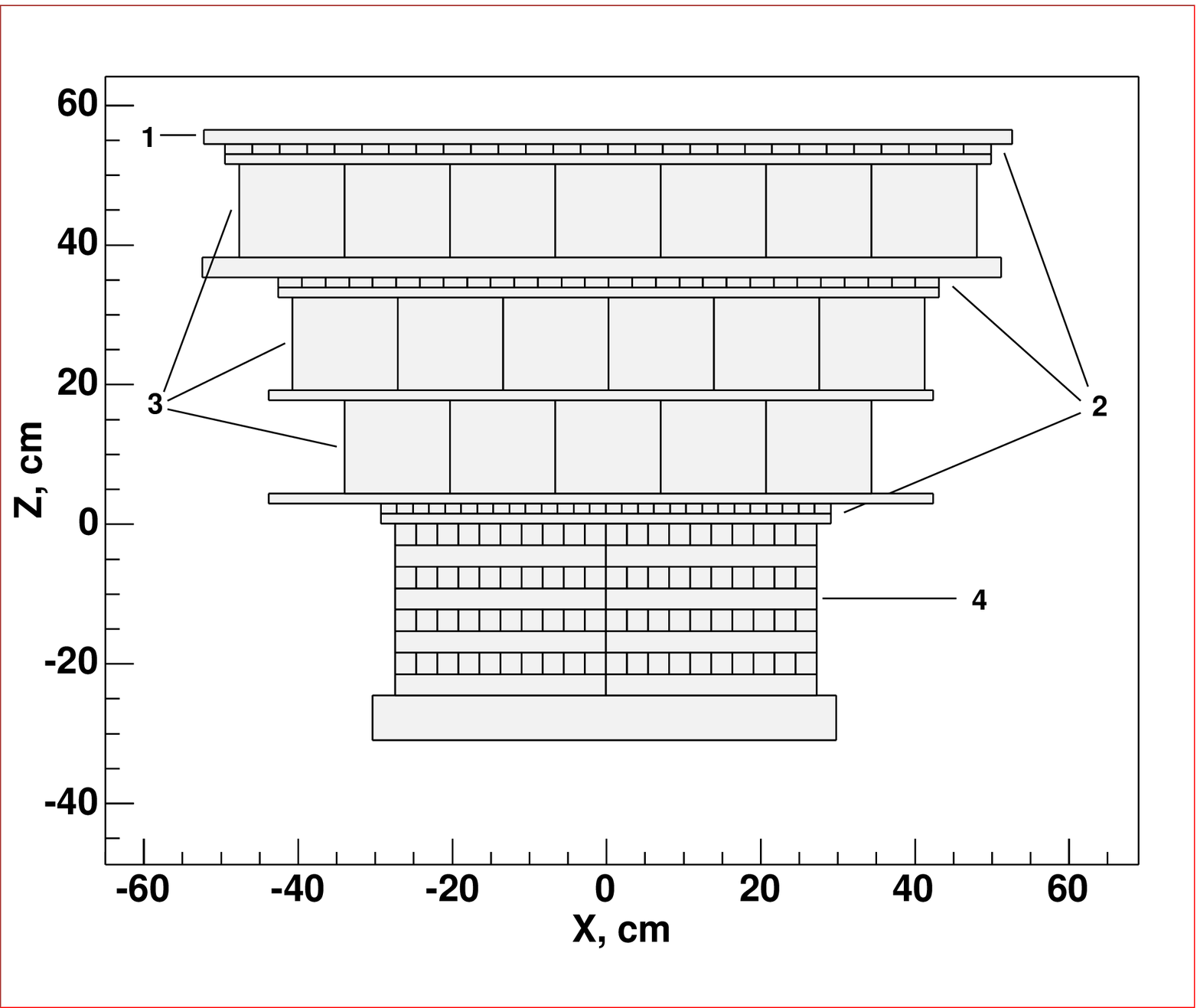}
\caption{\label{fig:Atic} 
The ATIC spectrometer (\mbox{ATIC-2} configuration). 1--silicon matrix, 2--scintillator hodoscopes, 3--carbon target, 4--BGO calorimeter.
}
\end{minipage}\hfill
\begin{minipage}[t]{\htw}
\begin{center}
\includegraphics[width=6.636cm]{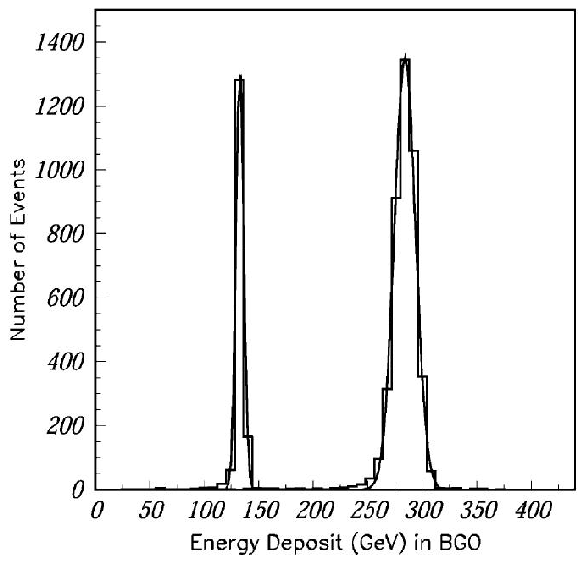}
\end{center}
\caption{\label{fig:ATIC-Resolution} 
Total BGO energy deposit for 150 GeV (left) and 300 GeV (right) electrons obtained in the beam tests \cite{ATIC-2005-GANEL-NIM}.
}
\end{minipage}\hspace{2pc}%
\end{figure}


The ATIC apparatus is comprised of a fully active bismuth germanate (BGO) calorimeter; a carbon target with embedded scintillator hodoscopes; and a silicon matrix that is used as a main charge detector as it shown in \fig{Atic}. 
The ATIC project had three successful flights around the South Pole in 2000--2001 (ATIC-1), 2002-2003 (ATIC-2) and 2007-2008 (ATIC-4). 
ATIC-1 was a test flight and is not discussed here. 
The details of construction of the apparatus and the procedures of its calibration are given in  \cite{ATIC-2004-GUZIK-AdvSpRes,ATIC-2004-ZATSEPIN-NIM,ATIC-2005-GANEL-NIM,ATIC-2008-PANOV-IET-ENG}.


The calorimeter of ATIC was thick for electrons. 
It was $18 X_0$ ($X_0$ is a radiative unit) in ATIC-2 and $22X_0$ in ATIC-4. 
As a result, the energy resolution for electrons of the instrument was high in the total energy range of 30~GeV--1~TeV.
The simulation predicted and the beam tests confirmed that the resolution, in terms of half width on the half height, was not worse than 3\% \cite{ATIC-2005-GANEL-NIM} (see \fig{ATIC-Resolution}).
Such high resolution, together with relatively high statistics, allows a meaningful measurement of the electron spectrum in the high-resolution mode -- with energy bins as small as 5--8\% (0.020--0.035 of decimal logarithm of energy).

\begin{figure}
\begin{minipage}[t]{\htw}
\includegraphics[width=\pictsize]{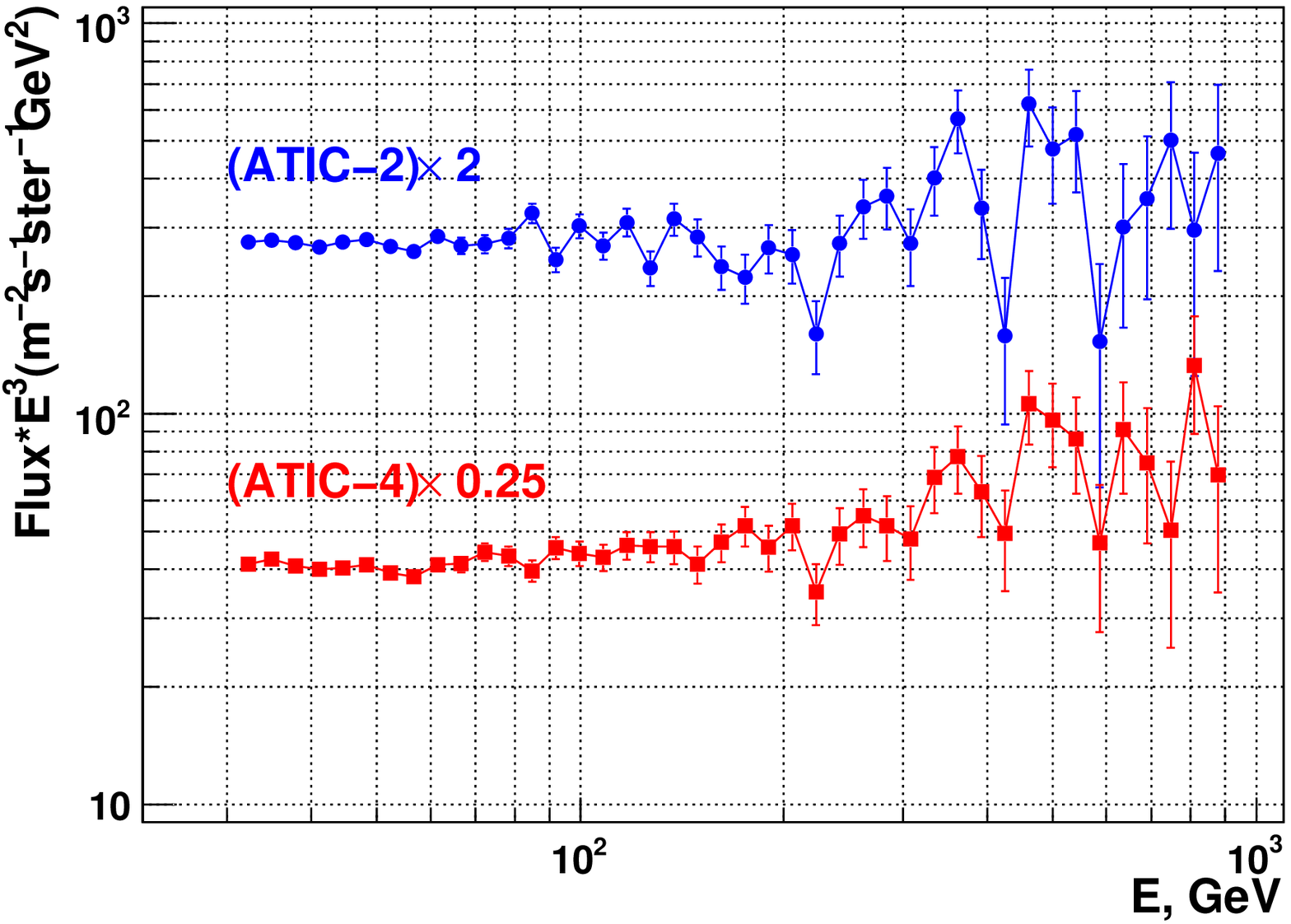}
\caption{\label{fig:A2-A4-Fine} 
The spectrum of electrons without the subtraction of proton background and without an atmospheric correction as measured in the ATIC-2 and ATIC-4 experiments \cite{ATIC-2011-PANOV-ASTRA}. The size of energy bin is 0.035 of the decimal logarithm of energy ($8.4\%$).
}
\end{minipage}\hfill
\begin{minipage}[t]{\htw}
\includegraphics[width=\pictsize]{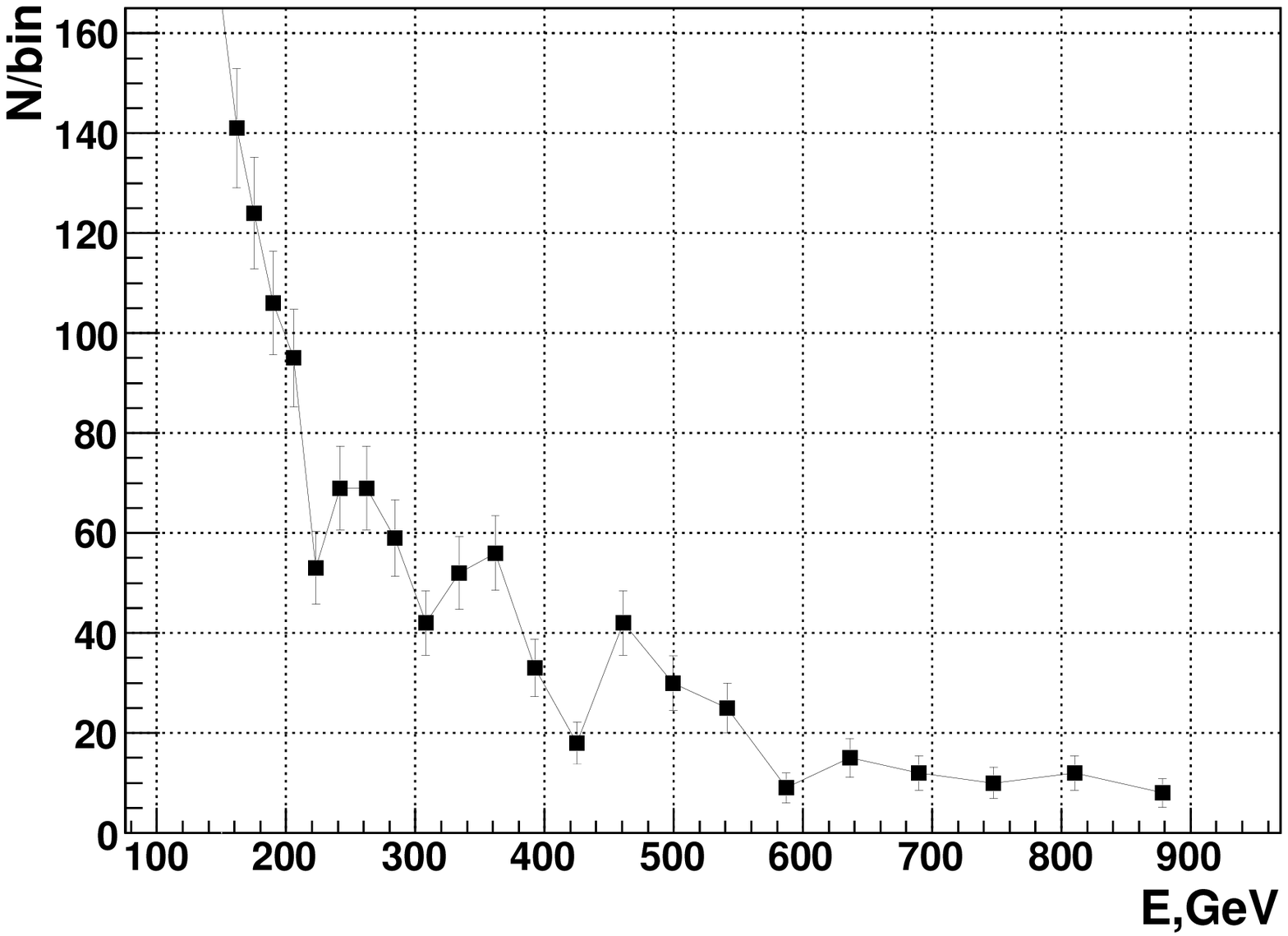}
\caption{\label{fig:N-A2A4-Fine} 
Total statistics of the electron spectrum in ATIC-2$+$ATIC-4 flights in each energy bin.  In this Figure, the summary spectrum corresponds to the spectra in \fig{A2-A4-Fine}.  In the range of 200-810 GeV, there were 701 events studied for statistical significance of the structure.
}
\end{minipage}\hspace{2pc}%
\end{figure}

Such high-resolution measurements were carried out for ATIC-2 and ATIC-4 flights in \cite{ATIC-2011-PANOV-ASTRA} and revealed a fine structure in the region of ``ATIC bump''(200--600~GeV) which was reproduced in both flights, see \fig{A2-A4-Fine}.
It can be seen that there were three peaks near 250~GeV, 350~GeV and 450~GeV respectively and three dips near 220~GeV, 300~GeV and 420~GeV.
\fig{A2-A4-Fine} shows the spectra of electrons without subtraction of the residual proton background and without an atmospheric correction.
Both the proton background and the factor of atmospheric correction are smooth functions and can not disturb the fine structure notably. 
Only these `raw' spectra are the input for the estimation of the statistical significance of the fine structure.
The statistical significance was estimated in \cite{ATIC-2011-PANOV-ASTRA} by two different ways. 
The first way was to use the usual $\chi^2$-test for the total ATIC-2+ATIC-4 spectrum. The second one was to use the degree of correlation between the structures measured separately by ATIC-2 and ATIC-4.
In both ways the statistical significance of the fine structure was estimated as 99.7\%. 
Please note that the bin statistics are not really low in the spectrum -- it varies from 70 events/bin to 40 events/bin near the three mentioned above peaks of the fine structure for the total spectrum ATIC-2+ATIC-4, see \fig{N-A2A4-Fine}.
The observed fine structure is expected to be stable against all known possible systematic errors since they cannot produce any short-scale structures in the spectrum.
Moreover, the observed fine structure passed and survived a number of tests on stability against possible systematics \cite{ATIC-2011-PANOV-ASTRA}.


Although the statistical significance of the observed phenomenon is not low and it looks systematically firm, it should be confirmed by independent experiments to be considered absolutely seriously of course.
However, if the structure is confirmed, it is very important in understanding the nature of the electron spectrum anomalies and the nature of DM.
As we explained in Section~\ref{SUBHALOS} above, the annihilation of DM particles could not produce several narrow peaks in the electron spectrum.
Therefore, DM annihilation could not be an origin of the observed fine structure.
Moreover, since the amplitude of the fine structure is very high, most of the intensity of the observed electron spectrum, at energies above 200~GeV, could not be attributed to DM annihilation or decay, therefore a high boost factor for DM annihilation is excluded.


The most natural explanation of the fine structure is a contribution of several nearby pulsars to the cosmic ray electron flux.
In fact, a fine structure of the electron spectrum which was very similar to the observed in the ATIC experiment, was predicted on the basis of nearby pulsars model in the article \cite{CRE-THEOR-MALYSHEV2009B} independently on the actual observation. 
The predicted fine structure is seen clearly in \fig{Malyshev-Pulsars}. 
Moreover, it was argued in \cite{CRE-THEOR-MALYSHEV2009B} that such features in the electron spectrum at high energies, would suggest strongly a pulsar origin of the anomalous contribution to the electron flux against its dark matter origin.
Let us note that the authors of the ATIC paper \cite{ATIC-2011-PANOV-ASTRA} observed the fine structure also independently of the paper \cite{CRE-THEOR-MALYSHEV2009B}. 
The first message about the fine structure was published in \cite{ATIC-2009-PANOV-ICRC} and the first preliminary spectra were shown in the paper \cite{ATIC-2010-PANOV-Lomonosov} without references to \cite{CRE-THEOR-MALYSHEV2009B}.


\begin{figure}
\begin{minipage}[t]{\htw}
\includegraphics[width=\pictsize]{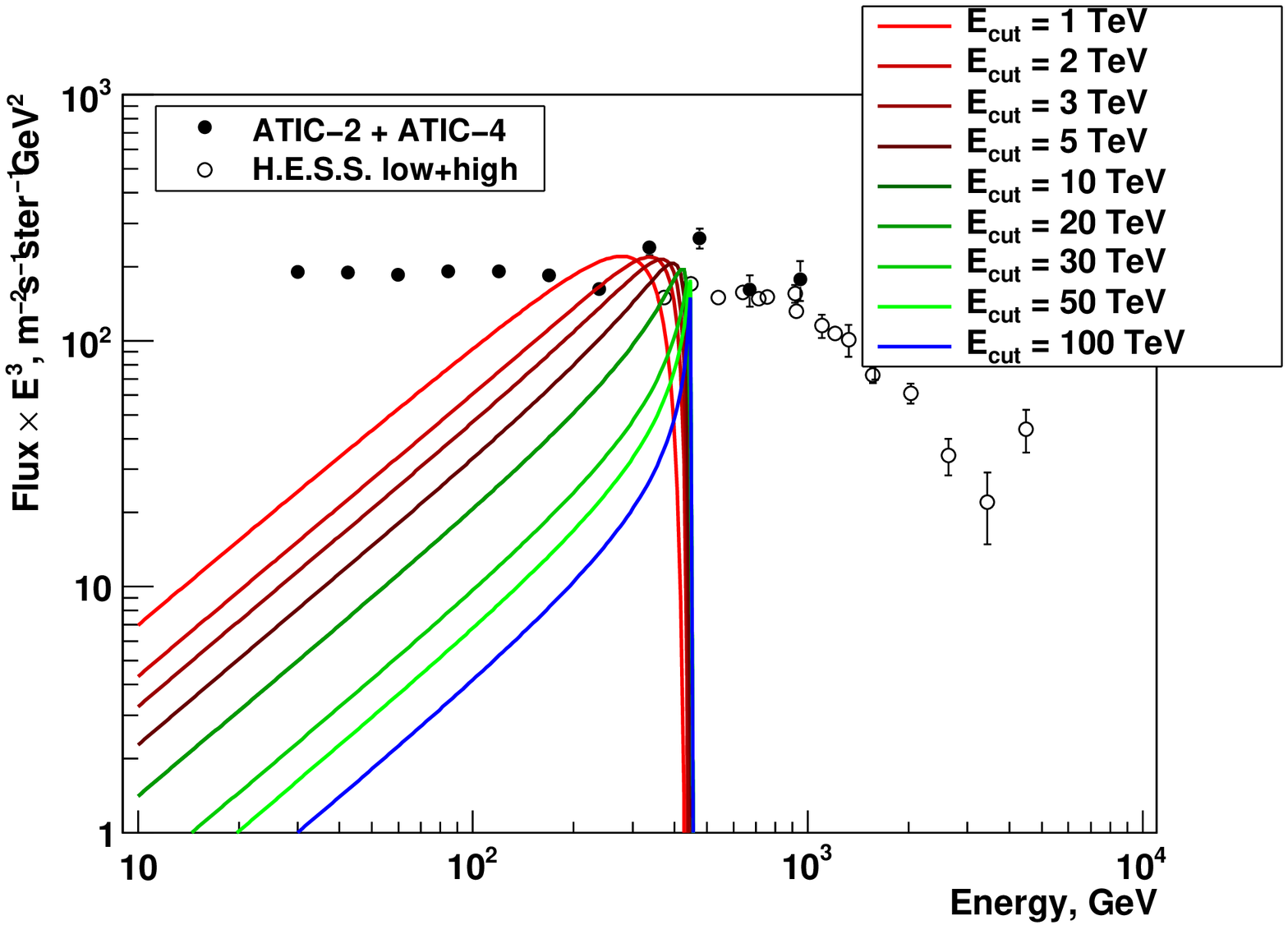}
\caption{\label{fig:Tr-ECut} 
A series of a single-pulsar electron spectra for different $E_{\mathrm{cut}}$ with other parameters 
$r=250$\,pc,
$\gamma=1.3$,
$\eta W_0=1\cdot10^{50}$\,erg,
$b_0=1.4\cdot10^{-16}$\,$(\mathrm{GeV}\cdot\mathrm{s})^{-1}$,
$D_0=3\cdot10^{28}$\,$\mathrm{cm}^2\cdot\mathrm{s}^{-1}$,
$\delta=0.3$.
}
\end{minipage}\hfill
\begin{minipage}[t]{\htw}
\includegraphics[width=\pictsize]{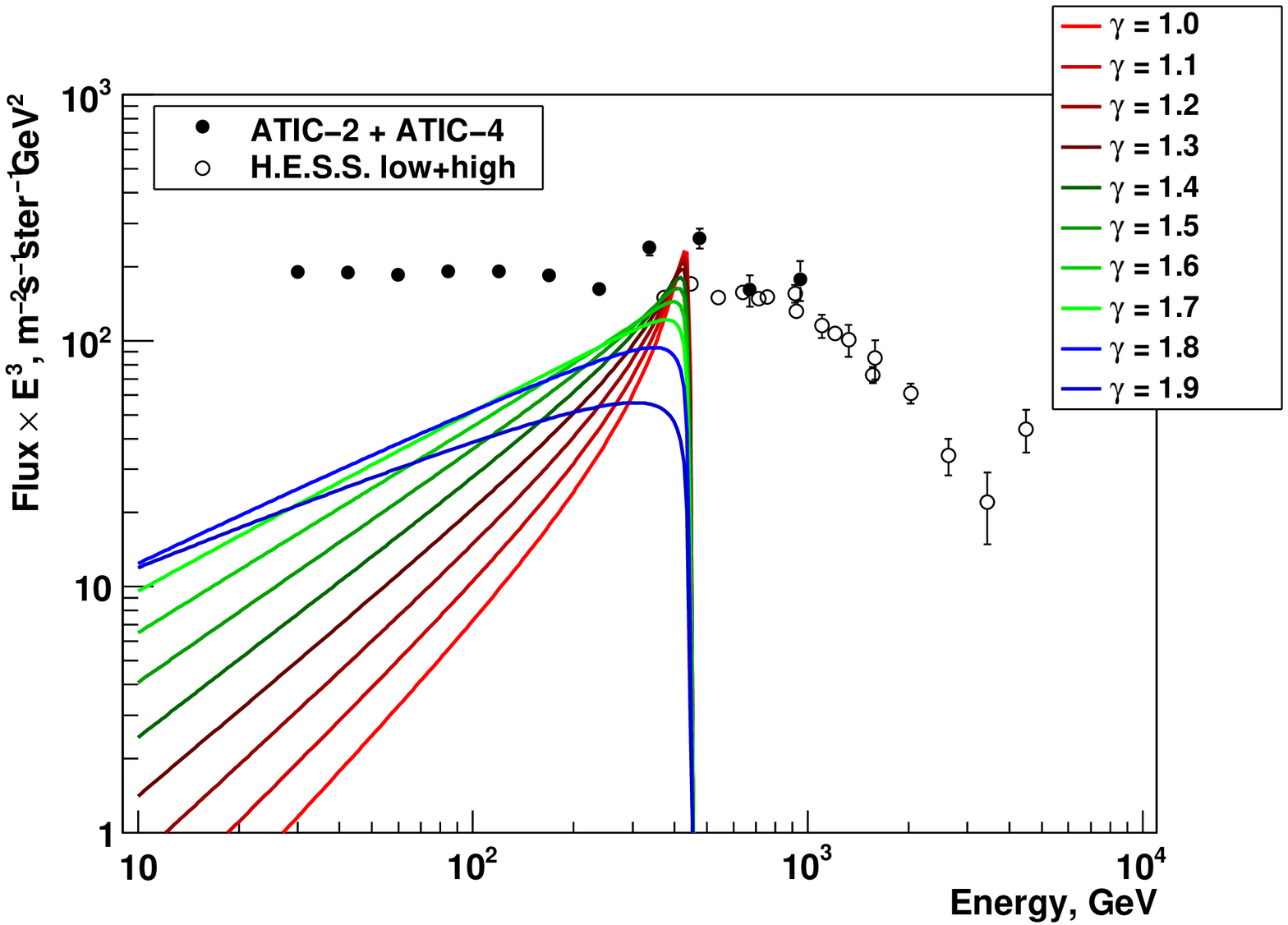}
\caption{\label{fig:Tr-Gamma} 
A series of a single-pulsar electron spectra for different source index $\gamma$ with other parameters 
$r=250$\,pc,
$E_{\mathrm{cut}}=10$\,TeV,
$\eta W_0=1\cdot10^{50}$\,erg,
$b_0=1.4\cdot10^{-16}$\,$(\mathrm{GeV}\cdot\mathrm{s})^{-1}$,
$D_0=3\cdot10^{28}$\,$\mathrm{cm}^2\cdot\mathrm{s}^{-1}$,
$\delta=0.3$.
}
\end{minipage}\hspace{2pc}\\%
\begin{minipage}[t]{\htw}
\includegraphics[width=\pictsize]{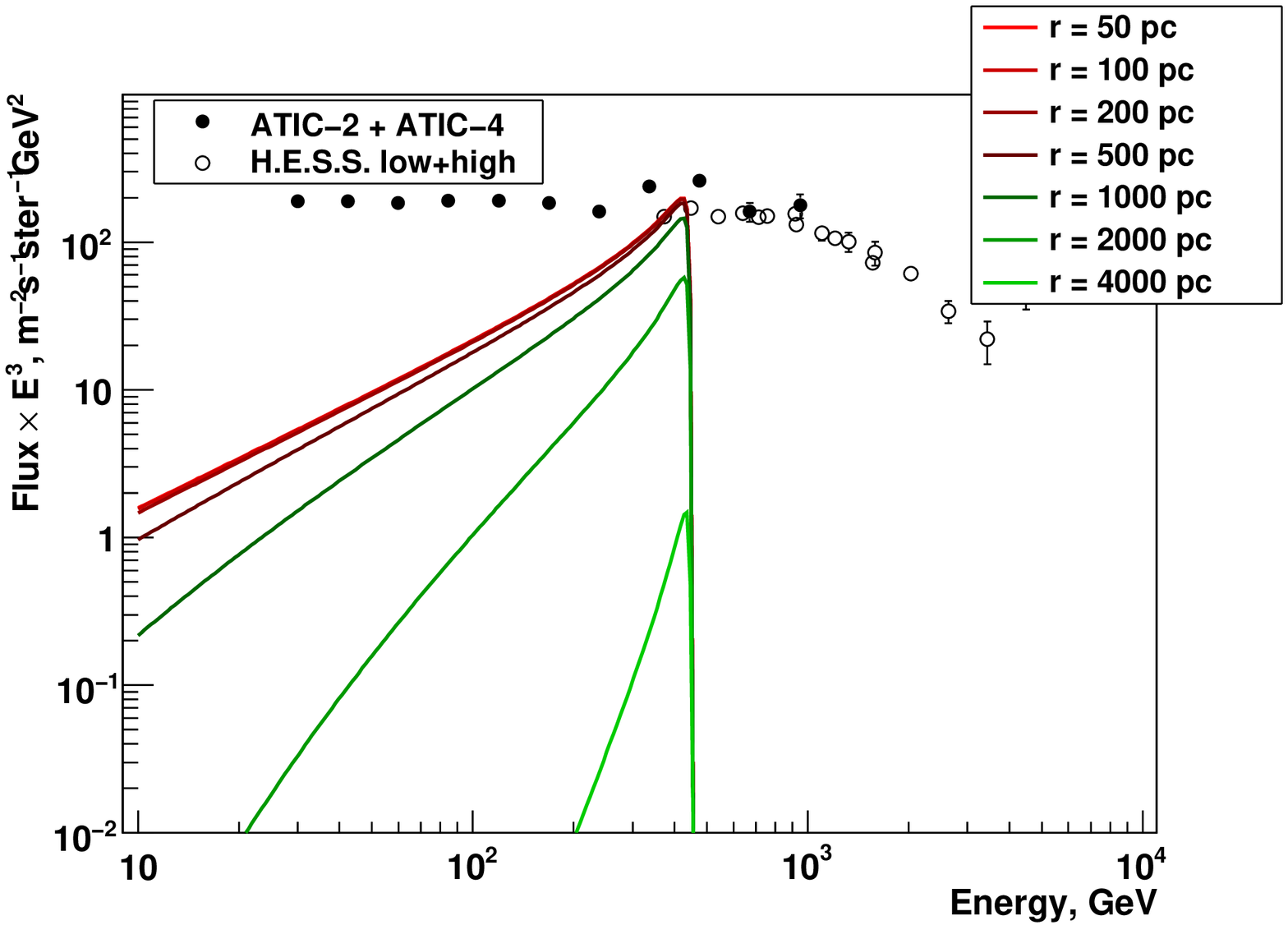}
\caption{\label{fig:Tr-R} 
A series of a single-pulsar electron spectra for different $r$ with other parameters 
$\gamma=1.3$,
$E_{\mathrm{cut}}=10$\,TeV,
$\eta W_0=1\cdot10^{50}$\,erg,
$b_0=1.4\cdot10^{-16}$\,$(\mathrm{GeV}\cdot\mathrm{s})^{-1}$,
$D_0=3\cdot10^{28}$\,$\mathrm{cm}^2\cdot\mathrm{s}^{-1}$,
$\delta=0.3$.
}
\end{minipage}\hfill
\begin{minipage}[t]{\htw}
\includegraphics[width=\pictsize]{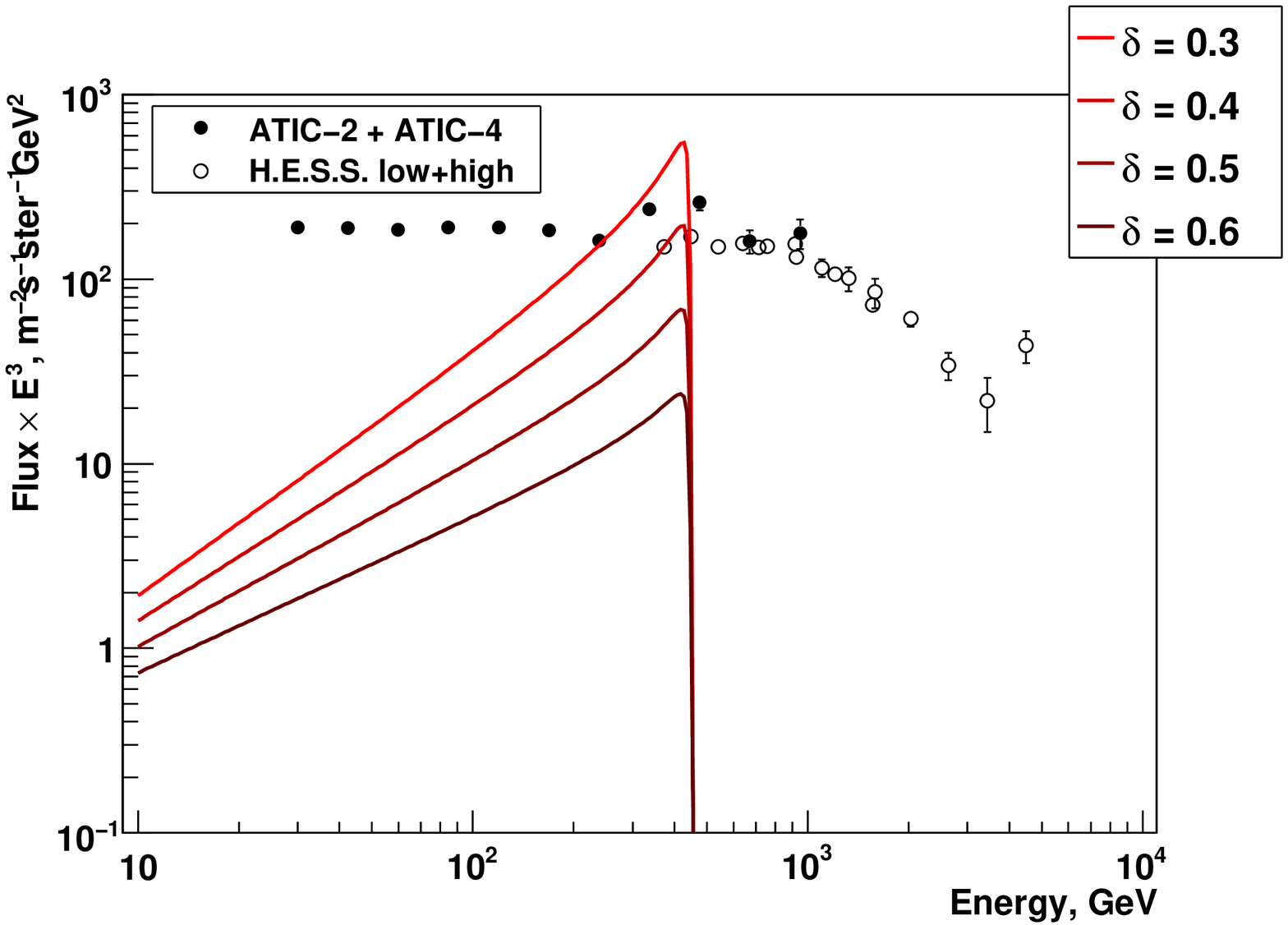}
\caption{\label{fig:Tr-Delta} 
A series of a single-pulsar electron spectra for different $\delta$ with other parameters 
$r=250$\,pc,
$\gamma=1.3$,
$E_{\mathrm{cut}}=10$\,TeV,
$\eta W_0=1\cdot10^{50}$\,erg,
$b_0=1.4\cdot10^{-16}$\,$(\mathrm{GeV}\cdot\mathrm{s})^{-1}$,
$D_0=3\cdot10^{28}$\,$\mathrm{cm}^2\cdot\mathrm{s}^{-1}$.
}
\end{minipage}\hspace{2pc}%
\end{figure}


\section{Fitting the fine structure by several single-pulsar spectra}


It can be seen from Figures  \ref{fig:A2-A4-Fine} and \ref{fig:N-A2A4-Fine} that the observed fine structure of the electron spectrum shows  an even higher amplitude than the structure predicted in \cite{CRE-THEOR-MALYSHEV2009B} (see \fig{Malyshev-Pulsars}). 
This poses a question: is it actually possible for nearby pulsars to produce a structure with so high an amplitude?
This question was discussed, for the first time, in the paper \cite{ATIC-2011-ZATSEPIN-ICRC}. 
There, it was argued that it  was possible, in principle; however, an unusual supposition had to be done about the cutoff of the conventional electron background near 200~GeV.
Now, we would like to discuss the same question, but supposing the usual power-low behaviour of the conventional background up to the energies of a few TeV.


\begin{figure}
\begin{minipage}[t]{\htw}
\includegraphics[width=\pictsize]{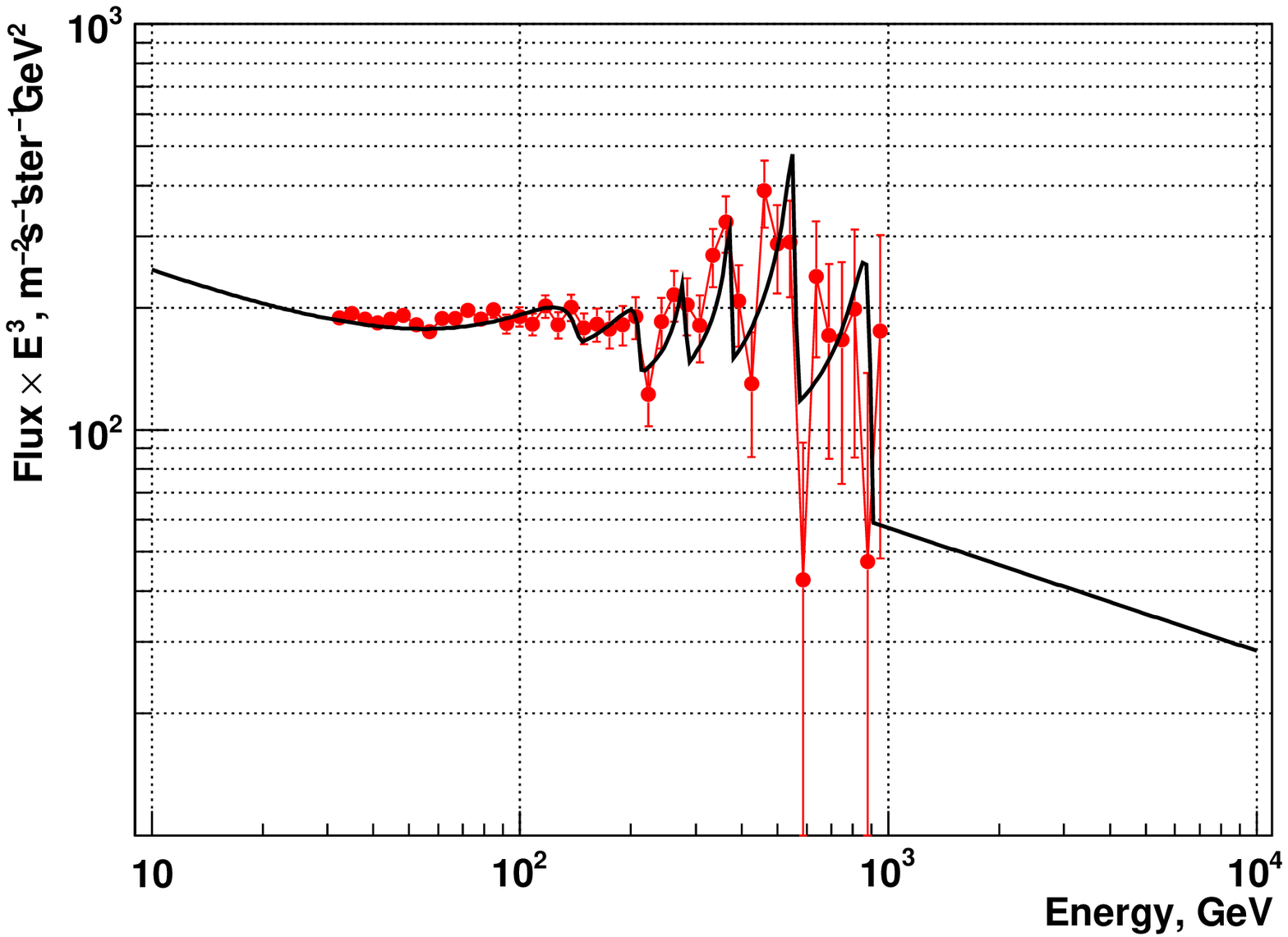}
\caption{\label{fig:ATIC-Fit-Electrons} 
An example of a fit of the measured fine structure in the electron spectrum by several nearby pulsars with conventional background. The papameters of the pulsars are shown in the Table~\ref{tab:PulsParam}, the propagation parameters are $\delta = 0.3$, $D_0 = 3\cdot10^{28}\,\mathrm{sm}^2\mathrm{s}^{-1}$, $b_0 = 1.4\cdot10^{-16}(\mathrm{GeV}\cdot\mathrm{s})^{-1}$.
}
\end{minipage}\hfill
\begin{minipage}[t]{\htw}
\includegraphics[width=\pictsize]{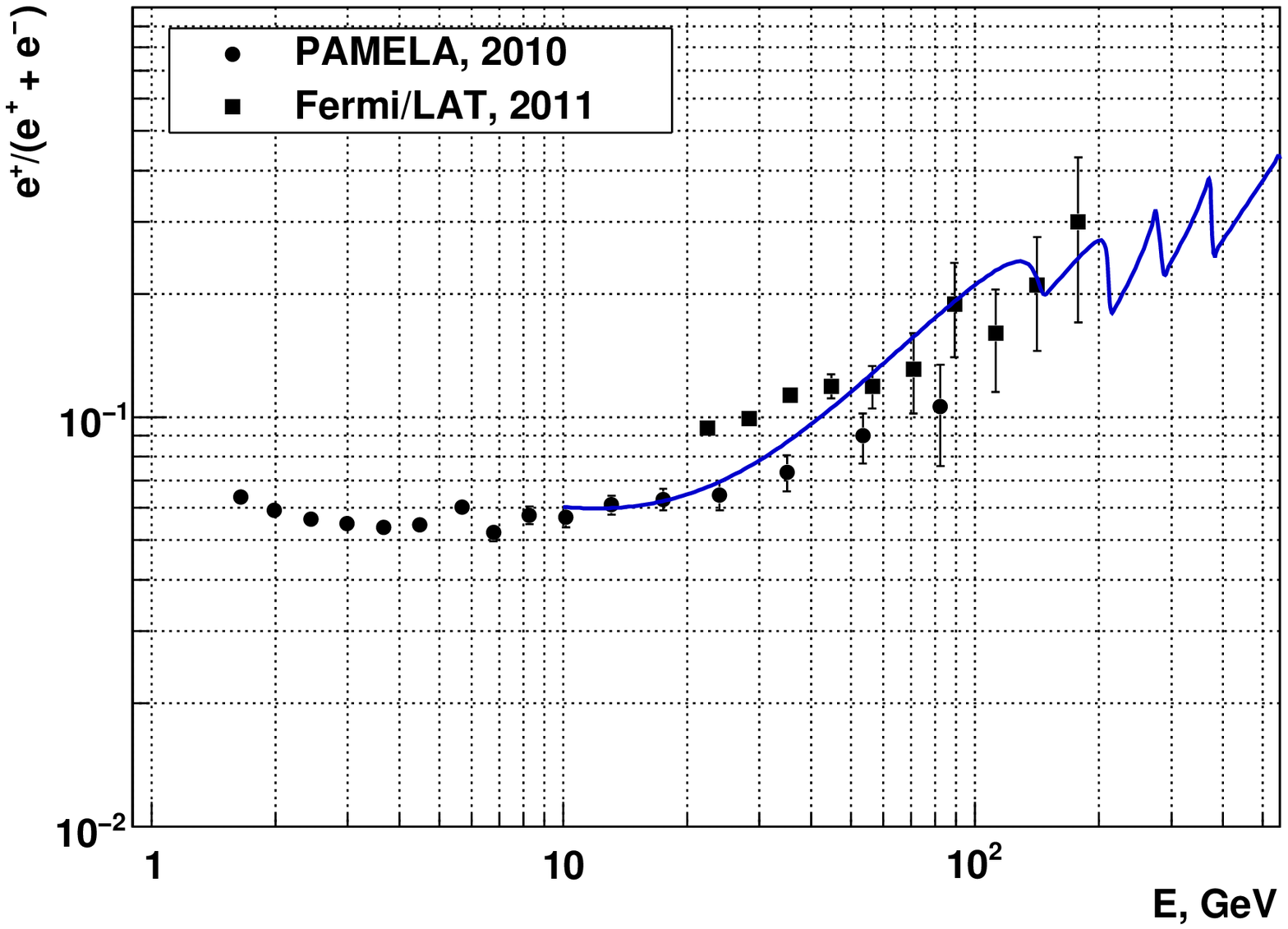}
\caption{\label{fig:ATIC-Fit-Positrons} 
The positron fraction calculated for the model of \fig{ATIC-Fit-Electrons}. The experimental data are: PAMELA, 2010 \cite{CRE-EXP-POS-PAMELA-2011-APh}; Fermi/LAT, 2011 \cite{CRE-EXP-POS-FERMI-2012-PRL}.
}
\end{minipage}\hspace{2pc}\\%
\end{figure}


In order to obtain a structure of the electron spectrum, like that observed in the ATIC experiment, obviously, the single-pulsar peaks, in the energy range 200--700~GeV, must be very sharp and narrow.
%
What may a single-pulsar spectrum look like?
Five series of single-pulsar electron spectra, for various parameters of pulsars and ISM, were obtained with the expression for a single-pulsar spectrum \eq{PulsarSolution}. These are shown in Figures \ref{fig:Tr-T-R1000ECut01gamma1.3}, \ref{fig:Tr-ECut}, \ref{fig:Tr-Gamma}, \ref{fig:Tr-R}, and \ref{fig:Tr-Delta}.
It can be seen from these Figures, that sharp and sufficiently strong single-pulsar electron peaks can be obtained from a variety ways with quite natural parameters.
By combining these ways, the measured fine structure can be fitted, also, to many different ways. 


Our task was not to identify each observed feature in the measured fine structure of the electron spectrum with some definite known pulsar.
Our opinion was that this problem was too difficult now  for two reasons.
Firstly, the distances to the nearby pulsars were known poorly (some other parameters, too, were known poorly).
%
Secondly, since the birth velocities of pulsars, due to asymmetric supernova explosions, were very high, the pulsars known locations had little in common with those locations where the pulsars had ejected the electrons.
The mean birth velocity was $450\pm90$~km/sec and a velocity above $1000$~km/sec was not rare \cite{PULS-LYNE1994}.
Typically, the difference between a current pulsar location and a location of burst-like ejection of the electrons by this pulsar may be hundreds of parsecs, and it is exactly the scale of distances from nearby pulsars to the Sun.
Therefore, the distances to actual bust-like sources are almost completely unknown and it is meaningless to try to fit the current positions of pulsars. Consequently, we tried to demonstrate only that the fit of the fine structure by pulsars was possible with some reasonable suppositions.


\fig{ATIC-Fit-Electrons} shows an example of a fit of the electron spectrum generated by several pulsars with the conventional background.
The conventional background is exactly the same as in figures \ref{fig:AllElectronsBefore2000}--\ref{fig:AllElectronsCurrentState}. 
The propagation parameters for this picture are $\delta = 0.3$, $D_0 = 3\cdot10^{28}\,\mathrm{sm}^2\mathrm{s}^{-1}$, $b_0 = 1.4\cdot10^{-16}(\mathrm{GeV}\cdot\mathrm{s})^{-1}$. 
The points in the figure denote the high-resolution electron spectrum measured by ATIC-2$+$ATIC-4 with the correct absolute normalization.
The proton background is subtracted; and the scattering of electrons in the residual atmosphere above the apparatus is corrected (see \cite{ATIC-2011-PANOV-ASTRA} for details).
The energy bin size is 0.035 of the decimal logarithm (8.4\%). 
Below 200~GeV, the data are fitted by two distinct pulsars; however, it should be considered only as an illustration.
Also, this region could be fitted by a continuum of far pulsars, as was done in paper \cite{CRE-THEOR-MALYSHEV2009B}.
The generated electron peaks is very sharp within the simple analytical model \eq{PulsarSolution}. 
Actually, the single-pulsar peaks are smeared out of course by the ISM inhomogeneity. However, this smearing is expected to be about only 5\% for $E\sim1$~TeV electrons \cite{CRE-THEOR-MALYSHEV2009B}.
Such a small smearing could not change notably the obtained picture.


\begin{table}
\caption{\label{tab:PulsParam} 
Parameters of pulsars used to generate \fig{ATIC-Fit-Electrons}.
}
 \begin{center}
  \begin{tabular}{|c|c|c|c|c|c|}
   \hline
   $nn$ & $r$,\,pc & age,\,kyr & $\gamma_{\mathrm{source}}$ & $E_{\mathrm{cut}}$,\,TeV & $\eta W_0$,\,erg \\
   \hline
   1 & 500 & 1500 & 1.5 &  1 & $1\cdot10^{50}$ \\
   2 & 500 & 1050 & 1.5 &  5 & $7\cdot10^{49}$ \\
   3 & 100 & 800  & 1.1 & 20 & $4\cdot10^{49}$ \\
   4 & 500 & 600  & 1.0 & 20 & $5\cdot10^{49}$ \\
   5 & 500 & 400  & 1.0 & 20 & $4\cdot10^{49}$ \\
   6 & 280 & 250  & 1.1 & 20 & $8\cdot10^{48}$ \\
   \hline
  \end{tabular}
 \end{center}
\end{table}

There are many other reasonable sets of parameters which, equally well, could fit the data -- this fit is a typical ill-defined problem.
However, it is important that, in any successful fit of the data, the fraction of the pulsars electron flux in the total flux of electrons above 200 GeV, must be very high to generate a structure with sufficiently high amplitude.
It suggests immediately that the positron fraction must rise very rapidly along the energy and must reach values as high as $\sim0.4$ near 400--500~GeV.
It is illustrated in \fig{ATIC-Fit-Positrons} that was generated by the same sets of parameters as \fig{ATIC-Fit-Electrons}. 
It is a very important and generic prediction of the model; we formulated it already in paper \cite{ATIC-2011-ZATSEPIN-ICRC}.
This prediction was done before the positron fraction, measured by Fermi/LAT \cite{CRE-EXP-POS-FERMI-2012-PRL}, was published, and, now, the model prediction was confirmed by the data up to the energies about 200 GeV. 

%
Anothr important prediction of the model is a fine structure that can be seen in the positron fraction curve at highest energies in \fig{ATIC-Fit-Positrons}.
It is not an artifact of the model. It is a generic implication of a variation of the pulsars flux in relation to the smooth conventional background in the energy range of the observed fine structure 200-600~GeV.
In principle, this fine structure may be measured, also. However, since the amplitude of the structure in the positron fraction is less that the fine structure amplitude in the electron spectrum, it is a more difficult problem than the observation of the fine structure in the total electron spectrum.
Also, there is a need for high-resolution measurements of the positron fraction to observe the phenomenon.


\section{Conclusions}

As understood from this review, there was a great progress, over the last ten years, in measurements and understanding of electrons and positrons in cosmic rays.  However, there is an extremely urgent need for new precise and high-statistical experiments. 


The information from recent measurements of the total electron spectrum is very contradictory.
There are a group of experiments which show a bump-like structure in the region 200--700~GeV: PPB-BETS \cite{CRE-EXP-PPB-BETS2008}, ATIC \cite{ATIC-2008-CHANG-NATURE}, ECC \cite{CRE-EXP-NISHIMURA2011-ICRC}, may be, also, the preliminary data of MAGIC \cite{CRE-EXP-MAGIC2011-arXiv}. 
Definitely, the Fermi/LAT spectrometer does not confirm the bump \cite{CRE-EXP-FERMILAT2010B}. The result of H.E.S.S. telescope is unclear in this respect, due to high energy threshold \cite{CRE-EXP-HESS2009}, and the PAMELA spectrum \cite{CRE-EXP-PAMELA2010-IzvFIAN} has too low statistics for definite conclusions in this region.
Also, the absolute flux differs for different modern experiments on a factor up to 3 in the region 300--500~GeV.


The positron fraction was measured by PAMELA \cite{CRE-EXP-POS-PAMELA-NATURE2009,CRE-EXP-PAMELA2011-PhysRevLett} and Fermi/LAT \cite{CRE-EXP-POS-FERMI-2011-1064,CRE-EXP-POS-FERMI-2011-ArXiv} up to 200~GeV only. 
Moreover, the data in the region 100--200~GeV are indirect since they were obtained with the help of the Earth Magnetic field, and not directly with a spectrometer.
There is an urgent need for high precision measurements of positron fraction for energies up to 500~GeV.


The fine structure of the electron spectrum, measured by high-resolution ATIC experiment \cite{ATIC-2011-PANOV-ASTRA}, could not be compared with any existing data since the ATIC spectrum was the only up to date high-resolution measurement of the electron spectrum.
The fine structure, measured by ATIC above 200~GeV, could not be resolved by using energy bins as wide as in the Fermi/LAT spectrum \cite{CRE-EXP-FERMILAT2010B} and in all other published data.
Moreover, the ATIC spectrum relates to the part of the Southern sky with the declination between $-45^{\mathrm o}$ and $-90^{\mathrm o}$, whilst the Fermi/LAT spectrum integrates the flux over the whole sky. 
Taking into account the possible anisotropy of the electron energy spectrum, it would be more correct to compare the ATIC spectrum with the spectrum measured by some high-resolution and high-statistical experiment for the same part of the sky.


The above noted lack of data is especially dramatic since a fine structure in the electron spectrum (and, also, in the positron fraction) may be a critical test  in selecting dark matter or local sources like nearby pulsars as an origin of anomalies in the electron spectrum and positron fraction (see Section \ref{FINESTRUCTURE}).
Usual experiments, with large energy binning in the spectrum, could not distinguish between dark matter and nearby pulsars as sources of cosmic ray electrons.
Therefore, there is a need not only for high-statistical and safe against systematic errors measurements but, also, for high-resolution measurements -- measurements with narrow energy bins ( $\approx$8\% or less). 


The author is grateful for discussion with D. Malyshev, N.V. Sokolskaya, V.I. Zatsepin. The work was supported by RFBR grant number 11-02-00275. 

\section*{References}

\providecommand{\newblock}{}

\end{document}